\documentclass[a4paper,twoside,11pt]{article}

\pdfoutput=1
\usepackage{cite,tabularx,cancel,xspace,flafter,ifpdf}

\usepackage[T1]{fontenc}
\usepackage{graphicx, caption, subcaption, color}
\usepackage{amsmath, amssymb, epstopdf, slashed}
\usepackage{parskip}
\usepackage{float}

\newcommand{\bk}[3]{
	\langle #1 | #2 | #3 \rangle
}

\newcommand{\gu}[1]{\gamma^{{#1}}}
\newcommand{\gl}[1]{\gamma_{{#1}}}
\newcommand{\zg}[0]{\ensuremath{Z/\gamma^*}}
\newcommand{\wbf}{VBF\xspace}
\newcommand{\hej}{High Energy Jets\xspace}
\newcommand{\Ca}{\ensuremath{C_{\!A}}\xspace}

\graphicspath{{Diagrams/}}

\title{\begin{normalsize}
\begin{flushright}
Edinburgh 2016/03\\
IPPP/16/19, DCPT/16/38\\
MCnet-16-8
\end{flushright}
\end{normalsize}
\vspace*{2.5cm} $\zg$ plus Multiple Hard Jets in High Energy Collisions
}

\author{Jeppe R.~Andersen$^{a}$, Jack J.~Medley$^b$,  Jennifer M.~Smillie$^b$\\ \mbox{}\\
  $^a$ Institute for Particle Physics Phenomenology,\\University of Durham,
  Durham DH1 3LE, U.K.\\
  $^b$ Higgs Centre for Theoretical Physics,\\University of Edinburgh,
  Edinburgh EH9 3FD, U.K.
}

\begin{document}
\maketitle

\begin{abstract}
  We present a description of the production of di-lepton pair
  production (through $Z$ boson and virtual photon) in association with at
  least two jets. This calculation adds to the fixed-order accuracy the dominant
  logarithms in the limit of large partonic centre-of-mass energy to all orders in the strong coupling $\alpha_s$.  This is achieved within the
  framework of High Energy Jets. This calculation is made possible by
  extending the high energy treatment to take into account the multiple
  $t$-channel exchanges arising from $Z$ and $\gamma^*$-emissions off several
  quark lines. The correct description of the interference effects from the
  various $t$-channel exchanges requires an extension of the subtraction
  terms in the all-order calculation.  We describe this construction and compare the resulting
  predictions to a number of recent analyses of LHC data. The description of
  a wide range of observables is good, and, as expected, stands out from other
  approaches in particular in the regions of large dijet invariant mass and large dijet rapidity spans.
\end{abstract}

\newpage
\tableofcontents

\section{Introduction}
\label{sec:Intro}

The Large Hadron Collider (LHC) sheds ever more light on Standard Model
processes at higher energies as it continues into Run II.  One
``standard candle'' process for the validation of  the
Standard Model description in this new energy regime is the production of a
dilepton pair through an intermediate $Z$ boson or photon, in
association with (at least) two jets
\cite{Chatrchyan:2011ne, Aad:2011qv,
  Chatrchyan:2013tna, Aad:2013ysa, Khachatryan:2014zya, Aad:2014rta,
  Khachatryan:2014dea}.  This final state can be entirely reconstructed from
visible particles (in contrast to $pp\to \text{dijets plus}(W\to) e\nu$) making it a particularly
clean channel for studying QCD radiation in the presence of a boson.
Experimentally, this process is indistinguishable from the production of a
virtual photon which has decayed into the same products, and we will consider
both throughout.

$W$ and \zg-production are excellent benchmark processes for investigating
QCD corrections, since the mass of the boson provides a perturbative scale,
while the event rates allow for jet selection criteria similar to those
applied in Higgs boson studies. $W,\zg$-production in association with dijets
is of particular interest, since in many respects it behaves like a dijet
production emitting a weak boson (i.e. electroweak corrections to a QCD
process rather than QCD corrections to a weak process). This observation
means that a study of $W,\zg$-production in association with dijets is
relevant for understanding Higgs-boson production in association with dijets
(which in the gluon-fusion channel can be viewed as a Higgs-boson correction
to dijet production). This process is interesting (e.g.~for $CP$-studies) in
the region of phase space with large dijet invariant mass, where the
coefficients in the perturbative series have logarithmically large
contributions to all orders. As an example of the increasing importance of
the higher orders, it is noted that the experimental measurement of the
$(N+1)/N$-jet rate in \zg+jets increases from 0.2 to 0.3 after application of
very modest \wbf-style selection cuts even at 7~TeV~\cite{Chatrchyan:2011ne,Aad:2011qv,Aad:2013ysa}.

The current state-of-the-art for fixed-order calculations for this process is the
next-to-leading order calculation of $\zg$ plus 4 jets by the BlackHat
collaboration~\cite{Ita:2011wn}.
While it has become standard to merge next-to-leading order QCD calculations
with parton
showers~\cite{Frixione:2002ik,Nason:2004rx,Frixione:2007vw,Alioli:2010xd,Frixione:2010wd,Alwall:2014hca},
results for jet production in association with $\zg$ bosons have so far only appeared with up to two
jets~\cite{Re:2012zi,Campbell:2013vha} (corresponding results for a $W$ boson
with up to three jets were given in~\cite{Hoeche:2012ft}, following those for a
$W$ boson plus two jets in~\cite{Frederix:2011ig,Campbell:2013vha}).  Indeed, $W/Z+0-$, $1-$ and
$2-$jet NLO samples have been merged with higher-multiplicity tree-level matrix
elements and parton
shower formulations~\cite{Hoeche:2012yf,Frederix:2015eii}. Beyond the
matching, the parton shower cannot be expected to accurately provide a
description of the large-invariant mass limit, from its resummation of
the (soft and collinear) logarithms which are enhanced in the region of small
invariant mass.  An alternative method to describe the higher-order
corrections is instead to sum the logarithmic corrections which are enhanced at
\emph{large} invariant mass between the particles.  This is the approach pioneered by
the High Energy Jets (HEJ) framework~\cite{Andersen:2009nu,Andersen:2009he}.
Here, the hard-scattering matrix elements for a given process are supplemented
with the leading-logarithmic corrections (in $s/t$) at all orders in
$\alpha_s$. This approach has been seen to give a good description of dijet and
$W$ plus dijet data at both the TeVatron~\cite{Abazov:2013gpa} and the
LHC~\cite{Aad:2011jz,Chatrchyan:2012gwa,Chatrchyan:2012pb,Aad:2014pua,Aad:2014qxa}.
In particular, these logarithmic corrections ensure a good description of $W$
plus dijet-production in the region of large invariant mass between the
two leading jets~\cite{Aad:2014qxa} and in large invariant mass regions in a
recent 4-jet ATLAS study~\cite{Aad:2015nda}. It is not surprising that standard
methods struggle in the region of large invariant mass, since the
perturbative coefficients receive large logarithmic corrections
to all orders, and perturbative stability is guaranteed only once these are systematically summed.

The purpose of this paper is to develop the treatment of such large QCD
perturbative corrections within \hej to include the process of \zg\ plus
dijets. While this process has many features in
common with the $W$ plus dijets process, one major difference is the
importance of interference terms, both between different diagrams within the
same subprocess (e.g. $qQ\to qQ(Z\to)e^+e^-$ with emissions off either the $q$ or
$Q$ line) and between $Z$ and $\gamma^*$
processes of the same partonic configuration. For processes with two quark
lines, the possibility to emit the $\zg$ from both of these leads to profound
differences to the formalism, since the $t$-channel momentum exchanged
between the two quark lines obviously differs depending on whether the boson emission is
off line $q$ or $Q$. Furthermore, the interference between the two resulting
amplitudes necessitates a treatment at the amplitude-level. \hej is
formulated
 at the amplitude-level, which, together with the matching to
 high-multiplicity matrix-elements, sets it apart in the field of
high energy
logarithms\cite{Kuraev:1976ge,Balitsky:1978ic,Lonnblad:1992tz,Lavesson:2005xu,Jung:2000hk,Jung:2010si,Colferai:2010wu,Caporale:2012ih,Ducloue:2012bm}. The
added complication over the earlier \hej-formalism (and indeed in any
BFKL-related study) by the interfering $t$-channels introduces a new structure of divergences in both real and virtual
corrections, and therefore a new set of subtraction terms are needed, in
order to
organise the cancellation of these divergences. The matching to full
high-multiplicity matrix elements puts the final result much closer to those
of fixed order samples merged according to the shower
formalism~\cite{Re:2012zi,Campbell:2013vha,Hoeche:2012yf,Frederix:2015eii}
--- although of course the logarithms systematically controlled with \hej are
different to those controlled in the parton shower formalism. In particular,
\hej remains a partonic generator, i.e.~although it is an all-order
calculation (like a parton shower), it is not interfaced to a hadronisation
model. Initial steps in combining the formalism of \hej and that of a parton
shower (and hadronisation) were performed in Ref.\cite{Andersen:2011zd}.

We begin the main body of this article by outlining the construction of a
High Energy Jets amplitude and its implementation in a fully flexible parton
level Monte Carlo in the next section.  In section~\ref{sec:Constructing} we
derive the new subtraction terms which allows us to fully account for
interference between the amplitudes. The subtraction terms allow for the
construction of the all-order contribution to the process as an explicit
phase-space integral over any number of emissions. Specifically, the
main result for the all-order summation is formulated in
Eq.~\eqref{eq:sigma}:
\begin{align}
  \nonumber
  \begin{split}
    \sigma =& \sum_{f_a,f_b} \sum_{n=2}^\infty  \left( \prod_{i=1}^n \int \frac{d^3 p_i}{(2\pi)^3
        2E_i} \right) \int \frac{d^3 p_{e^-}}{(2\pi)^3 2E_{e^-}} \int \frac{d^3 p_{e^+}}{(2\pi)^3 2E_{e^+}} \\
    & \ \times (2\pi)^4 \delta^{(2)}\left(\sum_i p_{i\perp} - p_{e^-\perp} -p_{e^+\perp}\right) \\
    & \ \times \ |\mathcal{M}^{HEJ-{\rm reg}}_{f_af_b\to \zg
      f_a(n-2)gf_b}(\{p_i\},p_{e^-},p_{e^+})|^2 \ \frac{ x_a f_{f_a}(x_a, Q_a) x_b
      f_{f_b}(x_b,Q_b)}{\hat s^2}\ \Theta_{\rm cut},
  \end{split}
\end{align}
where $\sigma$ is the sought-after cross
section, and the rest of the equation is discussed in the relevant section. Section~\ref{sec:Constructing} also discusses the necessary
modifications in order to include fixed-order matching. In
section~\ref{sec:Comparisons} we show and discuss the comparisons between the new
predictions obtained with \hej and LHC data. We conclude and present the
outlook in section~\ref{Conclusions}.


\section{The High Energy Limit of QCD and Real Corrections}
\label{sec:HEQCD}

Fadin and Lipatov observed\cite{Kuraev:1976ge,Balitsky:1978ic} that QCD scattering amplitudes
at large invariant mass (compared to the transverse momenta involved) exhibit the scaling
expected from Regge-theory. In particular, this means that for a given
configuration of the transverse momenta in a $2\to
n$-scattering, the limiting behaviour of the scattering amplitude as the
invariant mass between each pair of partons increases is dictated by the maximum spin
of any particle, which could be exchanged in what is termed the $t$-channel
between partons neighbouring in rapidity. This is found by ordering both
initial and final state particles according to rapidity (or light-cone
momenta in the case of incoming particles), and drawing all possible colour
connections between these. If a colour octet connection is allowed between
pairs of particles, this corresponds to the possibility of a spin-1 gluon
exchange, whereas colour-singlet exchange is identified as a spin-1/2 quark
exchange.

\begin{figure}
  \centering
  \includegraphics[width=0.95\textwidth]{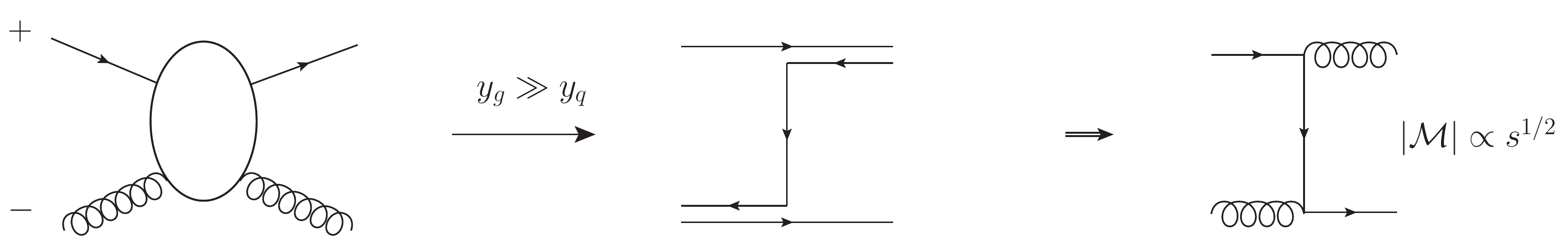}

  \vspace{0.3cm}
  \includegraphics[width=0.95\textwidth]{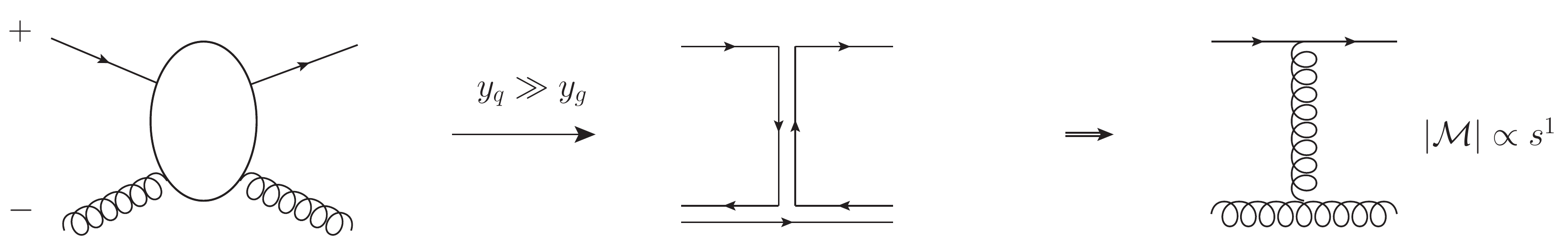}
  \caption{The two lines above illustrate the two possible rapidity orders for the
    process $qg\to qg$.  In the first case, where the rapidity of the gluon is
    greater than the quark, the allowed colour connection is a singlet
    corresponding to a quark exchange in the $t$-channel.  This leads to a
    contribution to the amplitude which scales as $s^{1/2}$.  In the second
    case, the allowed colour connection is an octet which corresponds to a gluon
  exchange in the $t$-channel and a scaling of $s^1$.  The latter will clearly
  be the dominant configuration in the limit of large $s$.}
  \label{fig:qgtoqg}
\end{figure}

The contribution to the cross section from a given momentum configuration of
the \emph{jets} (as opposed to partons) from the different flavour
assignments will have a different limiting behaviour, since the large
invariant-mass scaling is different e.g.~in the process of $qg\to qg$, if the rapidity
ordering of the final state $q$ and $g$ is swapped. Considering a specific
transverse momentum configuration of the jets in a simple $2\to 2$-process,
the full amplitude (which will then be squared in the calculation of the cross section) will scale as
$s^\omega$, where $s$ is the invariant mass of the final jets and $\omega$ is
the spin of the particle which would be exchanged in the $t$-channel.  Some cases, e.g.
$gg\to gg$, always allow for a gluon to  be exchanged, and
hence the amplitude scales as $s^1$ for large $s$.  In other cases,
e.g. $qg\to qg$, the $t$-channel particle exchanged is either a quark or
a gluon depending on the rapidity order of the flavour assignment, and hence the
amplitude scales as $s^{1/2}$ or
$s^1$ for large s. However, in this case, it is clear that in the limit of large $s$ the
contribution to the resulting jet momentum configuration will be dominated by
the process with the gluon exchange. This discussion is illustrated further
in Fig.~\ref{fig:qgtoqg}.
This argument may be further generalised to the case of more than two
outgoing partons, where now a $2\to n$ amplitude scales as
\begin{align}
  \label{eq:mrkscaling}
  |\mathcal{M}|\propto s_{12}^{\omega_1}\ \ldots\ s_{(n-1)n}^{\omega_{n-1}}\ \Gamma(\{t_i\}),
\end{align}
where the outgoing particles are ordered in rapidity, $s_{ij}$ is the
invariant mass of particles $i$ and $j$ and $\omega_i$ is the spin of the
particle exchanged in the $t$-channel of neighbouring particles. $
\Gamma(\{t_i\})$ depends only on the square of the $t$-channel
momenta (which in the limit corresponds to minus the square of their
transverse components).

We have thus identified the flavour-assignments of partons which will yield
the dominant contribution in the limit of large invariant mass between the
jets, for any given configuration of the transverse momenta: the dominant
contribution is obtained in the flavour configurations which allow for
colour-octet (gluon) exchanges between all neighbouring particles.  We call
these ``FKL configurations''.
Within \hej we concentrate on describing to all orders in the strong coupling
these scattering amplitudes, which contribute to the leading power behaviour of
the cross section.

These scaling arguments are unaffected by the additional emission of an electroweak boson
and specifically here we discuss the description with an additional $Z$ boson or
virtual photon.  The emission of an electroweak boson is viewed merely as an
electroweak correction to the underlying QCD dijet production.

\begin{figure}[btp]
  \centering
  \includegraphics[width=0.3\textwidth]{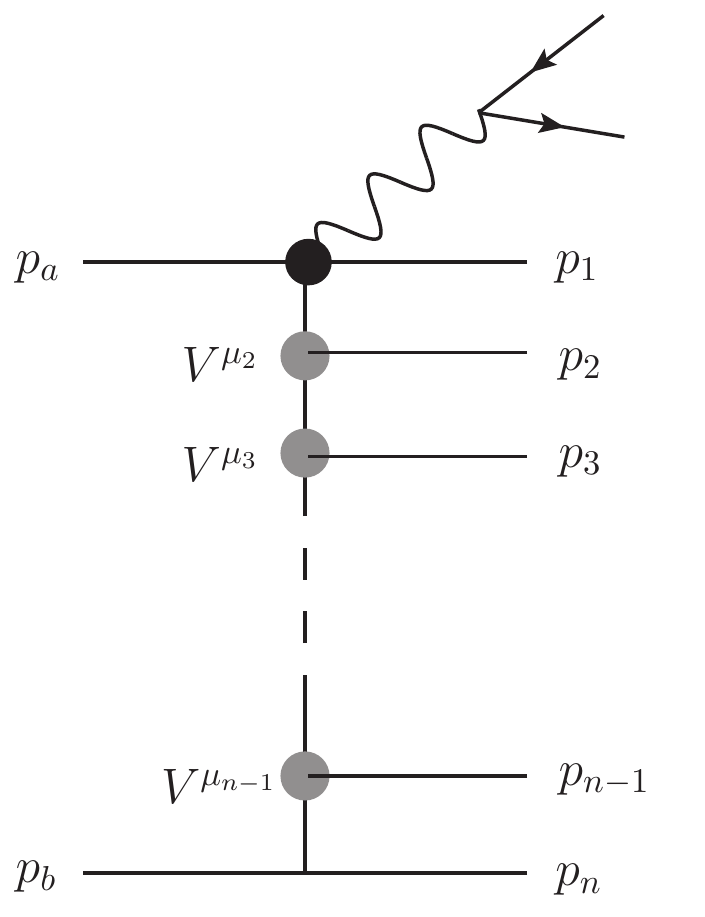}
  \caption{The schematic structure of the high-energy description of the matrix
    element for $qg\to qg...g(z\to)\ell^-\ell^+$, given in
    eq.~\eqref{eq:normal}.  In that specific case particles $a$ and 1 are quarks
    and particles $b$, 2,...,$n$ are gluons.}
  \label{fig:schema}
\end{figure}

We begin by considering $qg$-initiated processes where the quark is the
backward-moving incoming parton and take the leptonic decay of the $\zg$.  The
ordering described above motivates a unique definition of $t$-channel momenta,
namely if $p_a$ is the momentum of the backward quark, $p_b$ is the momentum of the forward gluon and
$y_1 \ll y_2 \ll ... \ll y_n$, one then defines $t_i = q_i^2$, where
$q_1=p_a-p_1-p_{\ell^+}-p_{\ell^-}$ and $q_i=q_{i-1}-p_i$ for $2\le i\le n$.
Furthermore, the leading contribution, which satisfies the requirement of maximal
$t$-channel gluon exchanges, arises purely from the outgoing state where all of the
intermediate particles in rapidity (those labelled 2 to $n-1$) must be gluons.
As discussed later, the factorisation property of amplitudes in the high-energy limit then allows us
to describe the emission of each of these gluons with an
independent effective emission vertex, a generalised Lipatov vertex $V^\mu$~\cite{Andersen:2009nu}, multiplying the corresponding
expression for the equivalent $2\to 2$ process, $qg\to qg(\zg\to)\ell^-\ell^+$ (see Fig.~\ref{fig:schema}).
At matrix-element-squared level
this gives
\begin{align}
  \label{eq:normal}
  \begin{split}
    \overline{|\mathcal{M}_{qg\to \zg qg..g}^{HE}|}^2 &=
    \overline{|\mathcal{M}_{qg\to \zg qg}^{HE}|}^2 \\ &\qquad \times
    \prod_{i=1}^{n-2} \left( g^2 \Ca \left( \frac{-1}{t_i t_{i+1}}
        V^\mu(q_i,q_{i+1}) V_\mu(q_i,q_{i+1}) \right)\right)
  \end{split}
\end{align}
where
\begin{align}
  \label{eq:Veff}
    \begin{split}
      V^\mu(q_i,q_{i+1})=&-(q_i+q_{i+1})^\mu \\
      &+ \frac{p_a^\mu}{2} \left( \frac{q_i^2}{p_{i+1}\cdot p_a} +
        \frac{p_{i+1}\cdot p_b}{p_a\cdot p_b} + \frac{p_{i+1}\cdot p_n}{p_a\cdot p_n}\right) +
      p_a \rightarrow p_1 \\
      &- \frac{p_b^\mu}{2} \left( \frac{q_{i+1}^2}{p_{i+1} \cdot p_b} + \frac{p_{i+1}\cdot
          p_a}{p_b\cdot p_a} + \frac{p_{i+1}\cdot p_1}{p_b\cdot p_1} \right) - p_b
      \rightarrow p_n.
    \end{split}
\end{align}
The lowest order expression on the right-hand-side of Eq.~\eqref{eq:normal},
$\overline{|\mathcal{M}_{qg\to \zg qg}^{HE}|}^2$, is the high-energy
description of the $q(p_a) g(p_b)\to q(p_1)g(p_n)(\zg \to)\ell^-(p_{\ell^-}) \ell^+(p_{\ell^+})$ process, which will be
described in full detail in section~\ref{sec:all-order-real}.  While $p_a+p_b
\ne p_{\ell^-}+p_{\ell^+}+p_1+p_n$ for $n>2$, the expression is built of
two independent factorised pieces, so this is not a problem.  Care needs to be
taken with the expression for the $t$-channel pole, which must be taken
symmetrically as $1/t^2=1/(t_1 t_{n-1})$.  
If the quark is instead the forward moving incoming parton, the expression is identical
except for the definition of $q_1$ where the lepton momenta is removed.

For
other initial states contributing to $\zg$ plus dijets, however, the situation
is more complicated. In particular for $qQ$-initiated processes, as the $\zg$
may be emitted from either quark line,
and there is interference from the two possibilities of exchanged $t$-channel momenta. The effective
emission vertex remains valid, but we must now work at amplitude level to
take into account this interference, both
here and for the virtual corrections as described in
section~\ref{sec:Constructing}.  In the remainder of this section we will
develop the equivalent of eq.~(\ref{eq:normal}) for all channels of $\zg$ plus
dijets.  We begin this in the next subsection, by describing our method of
constructing $\overline{|\mathcal{M}_{qg\to \zg qg}^{HE}|}^2$.

\subsection{Writing Matrix Elements in Terms of Currents}
\label{sec:factorising}

Traditionally, amplitudes in the HE limit are described as a product of two
scalar ``impact factors'', one for each end of the $t$-channel chain.  Instead,
in HEJ, we describe the core $2\to X+2$ processes in terms of a contraction of
two independent currents.  This is inspired by the structure of the exact
tree-level amplitudes, where each quark line automatically generates a current.
Effectively, helicity currents allow for the distinction of the kinematic
invariants $s$ and $u$, which is lost in the standard high-energy
factorisation at the cross-section level. This distinction proves
necessary in retaining accuracy in the approximations.  This can already be illustrated in the simple example of $qQ\to qQ$.
For all negative helicities for example, one can immediately write:
\begin{align}
  i\mathcal{M}_{q^-Q^-\rightarrow q^-Q^-} =  ig_s^2 T^d_{1a}T^d_{2b}\
  \frac{\bk{1}{\mu}{a}\cdot\bk{2}{\mu}{b}}{t},
  \label{eqn:jDefn}
\end{align}
where we have employed the spinor-helicity notation for the quark spinors, where
$\bk{i}{\mu}{j}$ is shorthand for $\bar u^-(p_i)\cancel{\gamma}^\mu u^-(p_j)$.
The repeated colour index $d$ is summed over and the lower colour indices refer
to their respective particle.

We will work in lightcone coordinates $p^\pm = E\pm p_z$ and further define
$p_\perp=p_x + i p_y$ and $e^{i\phi}=p_{\perp}/|p_{\perp}|$.  In components,
we get (using the spinors parametrised as in Ref.\cite{Andersen:2009nu})
\begin{align}
  \label{eq:cpts}
  i\mathcal{M}_{q^-Q^-\rightarrow q^-Q^-} =  ig_s^2 T^d_{1a}T^d_{2b}\
   \frac{2\sqrt{p_a^- p_b^+}}{t}
   \left( \sqrt{p_1^+ p_2^-} e^{i\phi_2} - \sqrt{p_1^- p_2^+} e^{i\phi_1} \right).
\end{align}

Let us first discuss the approach traditionally taken: in order to write this
in the desired factorised form of a product of scalars,
$C(p_a,p_1)\times C(p_b,p_2)$, it is necessary to use the limits
$p_1^+ \ll p_1^-$ and $p_2^- \ll p_2^+$ to neglect the first term.  If one
further approximates $p_1^- \simeq p_a^-$ and $p_b^+ \simeq p_2^+$, we may
write~\cite{DelDuca:1995zy}\footnote{Our spinor conventions differ by a phase to
  those in Ref.~\cite{DelDuca:1995zy} which vanishes in the matrix-element
  squared.}
\begin{align}
  i\mathcal{M}_{q^-Q^-\rightarrow q^-Q^-} = \frac{2s}{t} \left[  g_s T^d_{1a}
  e^{i\phi_1} \right] . \left[-i g_s T^d_{2b} \right].
\end{align}
This correctly captures the leading behaviour in $s/t$ and gives a factorised
expression.

However, by using helicity-currents, it is possible to achieve a form of
factorisation without relying on kinematic
approximations.  Returning to eq.~(\ref{eqn:jDefn}), it may immediately be
written as a contraction of two factorised four-vectors: $V(p_a,p_1).V(p_b,p_2)$,
where the vectors depend on the same momenta as the factorised vertices in
the traditional approach, but now the vectors (up to constants) are just standard currents
$j^{-\mu}(p_i,p_j)=\bk{i}{\mu}{j}$:
\begin{align}
  \label{eqn:jexpr}
  i\mathcal{M}_{q^-Q^-\rightarrow q^-Q^-} \equiv  ig_s^2 T^d_{1a}T^d_{2b}\ \frac{j_1^\mu\cdot
  j_{2\mu}}{t}.
\end{align}
Each helicity current has two independent components and this extra degree of
freedom compared to the impact factors of the traditional approach is
precisely what is required in order to keep the first term in eq.~(\ref{eq:cpts}) and
therefore describe the amplitude exactly.

This illustration is clearly for a very simple process, but the same conclusion
applies more generally.  One can exactly describe $qg\to qg$ as the contraction
of a standard  quark current and a gluon current $j_\mu^g$, consisting of a
product of a standard quark current and colour factors depending on
the gluon momenta only\cite{Andersen:2009he}. This holds even though the $qg$-scattering process has $s,t$
and $u$-singularities.  The same holds for $gg\to gg$ as long as the
helicities of the two incoming (and outgoing) gluons differ, such that one
can define the $s,t,u$-channels. One can also go beyond pure QCD and describe
$qQ\to W q'Q$, $qQ\to \zg qQ$ and $qQ\to qQH$ exactly as the contraction of two
currents~\cite{Andersen:2009nu}.  In the next subsection we describe the new
current for $\zg$ plus jets, and the construction of the resulting amplitude.

\subsection{A Current for $\zg$ plus Jets}
\label{sub:ZCurrents}

\begin{figure}[btp]
  \begin{center}
    \includegraphics[width=1.0\linewidth]{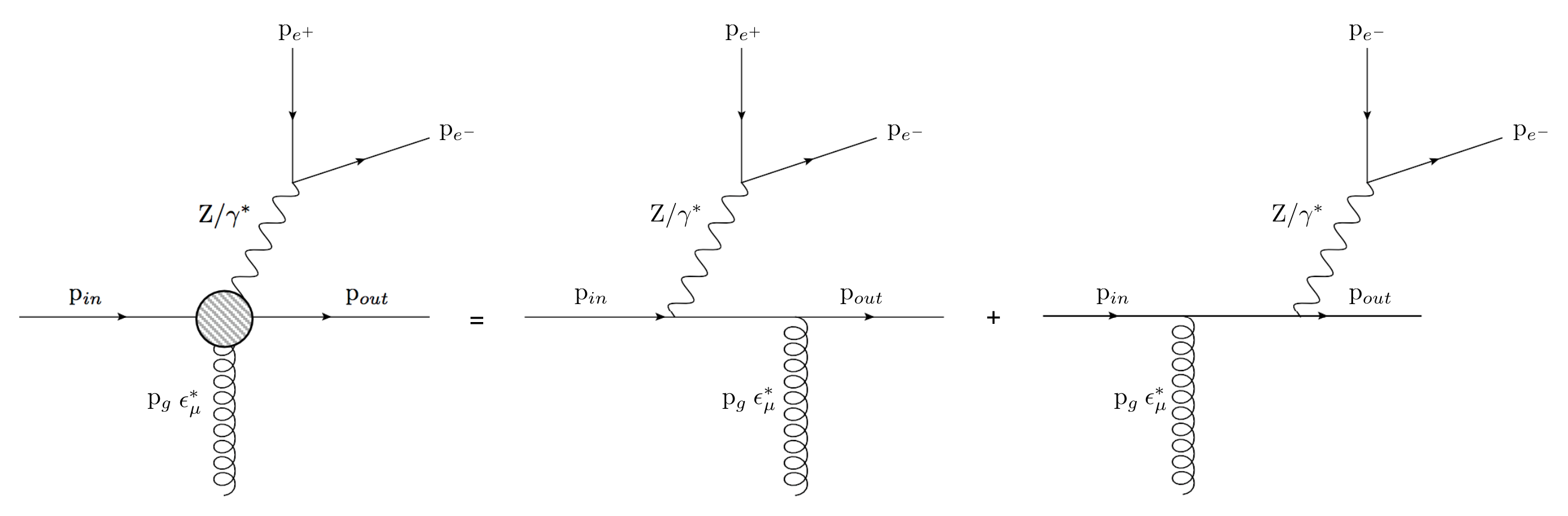}
    \caption{The current used to describe the quark line with the emission of a
      $Z$ or $\gamma^*$ is the sum of the contributions arising from the two possible emission sites for the $\zg$.}
    \label{fig:ZCurrent}
  \end{center}
\end{figure}

In this section, we will construct a current to describe the emission of a $\zg$
boson and exchange of a $t$-channel gluon from a quark or antiquark line.  
We can
write the current for the $Z$ emission (only), $j_Z^\mu$, as a sum of the
contributions from the two possible emission sites: one where the $Z$ is emitted
before the $t$-channel gluon and another where the gluon is radiated first,
shown diagramatically in figure~\ref{fig:ZCurrent}.  For definiteness, we
could then
consider the decay $Z\to e^+e^-$.  We have
\begin{align}
  \begin{split}
    j^Z_\mu = \frac{C_{Zq}C_{Ze}}{p_Z^2 - M^2_Z +
      i\Gamma_ZM_Z}\Big(&\frac{\bk{1}{\gu{\sigma}(\slashed p_{out} + \slashed
        p_{e^+} + \slashed p_{e^-})\gl{\mu}}{a}}{(p_{out} + p_Z)^2} + \\
    &\frac{\bk{1}{\gu{\mu}(\slashed p_{in} - \slashed p_{e^+} - \slashed
        p_{e^-})\gl{\sigma}}{a}}{(p_{in} - p_Z)^2}\Big)\bk{e^+}{\gl{\sigma}}{e^-},
    \label{eqn:Zcurrent}
  \end{split}
\end{align}
where $M_Z$ is the mass of the $Z$ boson, $\Gamma_Z$ is its width, $C_{Zx}$ is
the coupling of the $Z$ to $x$, $x=e,q,\nu_e,\ldots$ and $\mu$ is the Lorentz
index for the $t$-channel gluon propagator.
Expanding the quark and lepton momenta using their completeness relations we can
fix the helicity of the incoming quark, $h_{in}$, and the outgoing quark,
$h_{out}$, to be identical, and we are left with a current which only has four
possible helicity configurations depending on $h_q = h_{in} = h_{out}$ and the
electron helicity, $h_e$:
\begin{align}
  \begin{split}
    j^Z_\mu(h_q, h_e) = &\ C_{Zq}^{h_q}C_{Ze}^{h_e}\ \frac{\bk{e^+_{h_e}}{\gl{\sigma}}{e^-_{h_e}}}{p_Z^2 -
      M^2_Z + i\Gamma_ZM_Z} \\ &\ \times\bigg(\frac{2p_1^\sigma\bk{1_{h_q}}{\gu{\mu}}{a_{h_q}} +
      \bk{1_{h_q}}{\gu{\sigma}}{e^+_{h_q}} \bk{e^+_{h_q}}{\gu{\mu}}{a_{h_q}} +
      \bk{1_{h_q}}{\gu{\sigma}}{e^-_{h_q}} \bk{e^-_{h_q}}{\gu{\mu}}{a_{h_q}}}
    {(p_{out} + p_Z)^2}\\
    &\qquad + \frac{ 2p_a^\sigma\bk{1_{h_q}}{\gu{\mu}}{a_{h_q}} -
      \bk{1_{h_q}}{\gu{\mu}}{e^+_{h_q}} \bk{e^+_{h_q}}{\gu{\sigma}}{a_{h_q}} -
      \bk{1_{h_q}}{\gu{\mu}}{e^-_{h_q}} \bk{e^-_{h_q}}{\gu{\sigma}}{a_{h_q}}}
    {(p_{in} - p_Z)^2}\bigg).
    \label{eq:Zcurrent}
  \end{split}
\end{align}

For the charged lepton channels for $Z$-decays, we must also include the contribution arising from the exchange of an off-shell photon,
$\gamma^*$.  The expression for the current for the off-shell photon has the
same form to that shown in eq.~(\ref{eq:Zcurrent}) with the $Z$ propagator
replaced with that of the photon and the couplings modified.  Our final current
is then the sum of the two:
\begin{align}
  \label{eq:jsum}
  j^{\zg}_\mu(h_q,h_e) = j^Z_\mu(h_q,h_e) + j^\gamma_\mu(h_q,h_e).
\end{align}

\boldmath
\subsection{All-Order Real Corrections for $\zg$ Plus Dijets}
\unboldmath
\label{sec:all-order-real}

With the current derived in the previous subsection, we have the required
building blocks to describe the dominant contribution to the real emission in the HE limit,
in the manner of eq.~(\ref{eq:normal}).  We first construct the lowest order
description, $\overline{|\mathcal{M}_{qg\to Zqg}^{HE}|}^2$.  Our current,
$j^{\zg}_\mu(h_q,h_e)$, is already the sum of diagrams with a mediating $Z$ and
diagrams with a mediating $\gamma^*$.  For the quark-gluon initiated processes,
this is then all we need for the complete amplitude and we write:
\begin{align}
  \label{eq:qgamp}
  \begin{split}
    \overline{|\mathcal{M}_{qg\to Zqg}^{HE}|}^2 =& \ \frac{g_s^2}{8}\
    \frac{1}{(p_a-p_1-p_{e^+}-p_{e^-})^2 (p_b-p_n)^2}  \sum_{h_q,h_e,h_g} |
    j^{\zg}_\mu(h_q,h_e) j^{g\mu}(h_g) |^2.
  \end{split}
\end{align}
The interference term between the $Z$ and $\gamma^*$ processes is immediately
included in this construction through squaring the sum of Eq.~(\ref{eq:jsum}).  The equivalent
expressions for the $gq$-initial state and for $\bar{q} g$ and
$g\bar{q}$-initial states all have the same simple form.  This can then be
substituted into eq.~(\ref{eq:normal}) to give the real corrections up to any
order in $\alpha_s$.

\begin{figure}[btp]
  \begin{center}
    \includegraphics[width=0.9\linewidth]{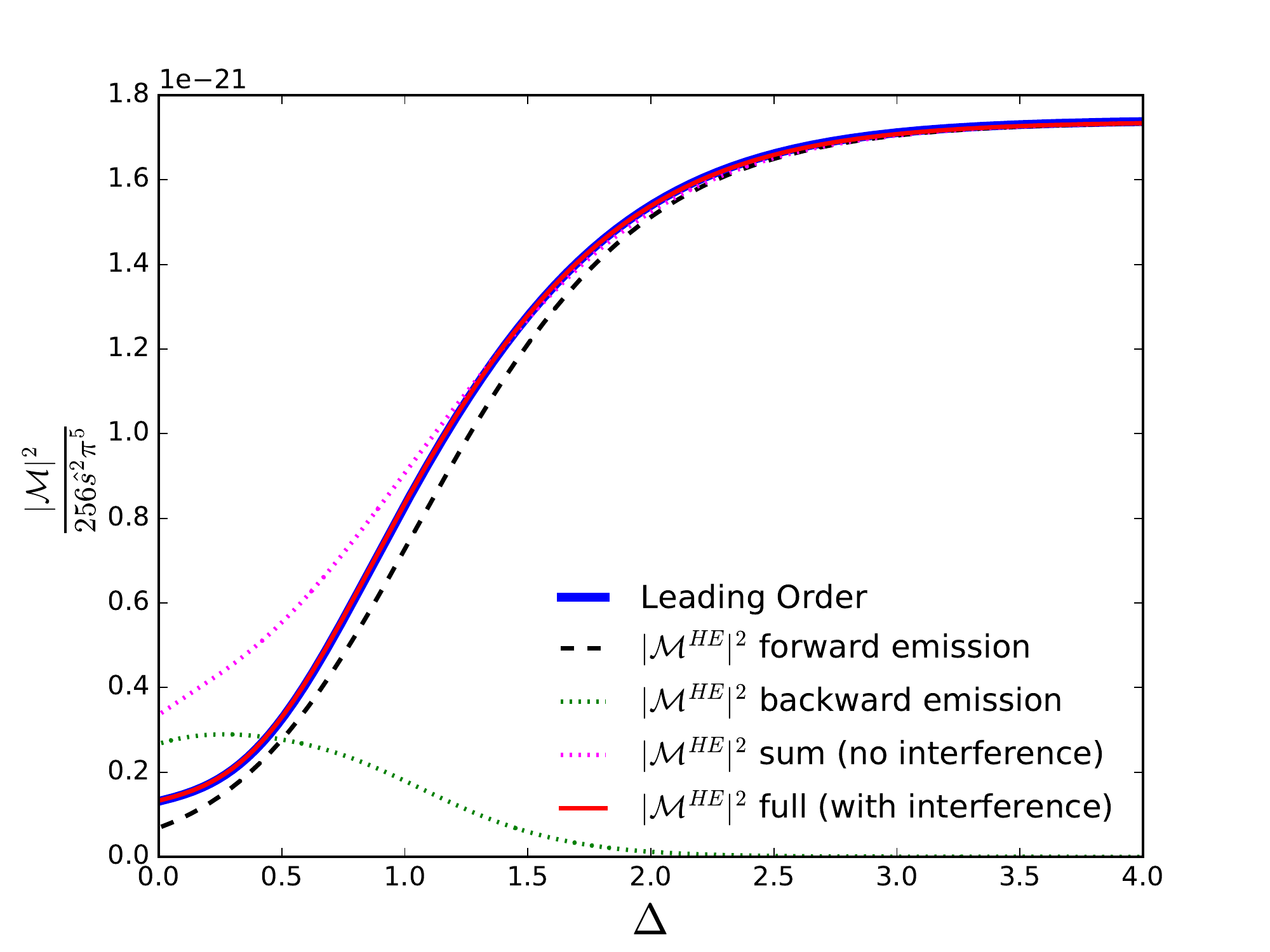}
    \caption{The matrix-element squared divided by the square of the partonic
      centre-of-mass energy for $qQ\to ZqQ$ with the $Z$ decaying to an
      electron-positron pair for the phase space slice described in
      eq.~\eqref{eq:momenta}.  Increasing values of $\Delta$ represent
      increasing rapidity separation between the jets.  The different lines show the contributions from
      different terms in the calculation: only emission from the forward or the
      backward quark line (black, dashed and green, dotted), their sum without
      the interference term (magenta, dotted) and their sum including
      interference (red, solid) which is seen to agree exactly with the LO result
      (blue, thick solid).}
    \label{fig:twojets}
  \end{center}
\end{figure}

We now turn our attention to the case of two incoming quark lines (or a mix
of quark and anti-quarks). Here, it
is possible for the $Z$ to be emitted from either quark line, and it turns
out that the interference effects are sizeable, see Fig.~\ref{fig:twojets}.  We must
include both possibilities and allow for the interference term.  Our high-energy
description of the matrix elements relies on the correct description of the
$t$-channel momenta, and this obviously depends on which of the quark lines
the $Z$ or $\gamma^*$ was emitted from.  We therefore need to modify the simple
framework outlined above.  We will use the subscript $a$ ($b$) to label the
current at the lowest (highest) end of the rapidity chain.  We then define $t_a$
($t_b$) to be the $t$-channel momentum exchanged when the bosons are emitted at
the lowest (highest) end of the rapidity chain.  Then the full amplitude squared
for $qQ\to qQ(\zg\to) e^+ e^-$ is given by:
\begin{align}
  \begin{split}
    \overline{|\mathcal{M}_{qQ\to ZqQ}^{HE}|}^2 &= g_s^2 \frac{C_F}{8N_c} \Big|\frac{j^{\zg}_a\cdot j_b}{t_a} + \frac{j_a\cdot
      j^{\zg}_b}{t_b}\Big|^2\\
    &= g_s^2 \frac{C_F}{8N_c} \left( \Big|\frac{j^{\zg}_a\cdot j_b}{t_a}\Big|^2 + \Big|\frac{j_a\cdot
      j^{\zg}_b}{t_b}\Big|^2 + 2\Re{\Big\{\Big(\frac{j^{\zg}_a\cdot
        j_b}{t_a}\Big)\Big(\frac{j_a\cdot j^{\zg}_b}{t_b}\Big)^*\Big\}} \right),
    \label{eqn:interference}
  \end{split}
\end{align}
where $j_{a,b}$ are the pure quark currents defined above eq.~(\ref{eqn:jexpr}).  The
coupling constants of the $Z$ to the relevant quarks and leptons are contained
within $j^{\zg}(h_q,h_e)$, as in eq.~(\ref{eqn:Zcurrent}).  Fig.~\ref{fig:twojets}
shows the value of this matrix element squared divided by the squared partonic
centre-of-mass energy for increasing rapidity separation of the two jets. The
result is compared with that obtained from the full, tree-level matrix elements from
MadGraph5\_aMC@NLO~\cite{Alwall:2014hca}.
The slice through phase space here is given by:
\begin{align}
  \label{eq:momenta}
  \begin{split}
    p_i = &(k_{i\perp} \cosh y_i; k_{i\perp} \cos \varphi_i, k_{i\perp} \sin \varphi_i, k_{i\perp}
    \sinh y_i) \\
    \mbox{with} \hspace{2.8cm} & \\
    k_{1\perp} = k_{e^+\perp} = 40\mbox{GeV}& \hspace{0.5cm} k_{e^-\perp} =
    \frac{m_Z^2}{2k_{e^+\perp}\left(\cosh(y_{e^+} - y_{e^-}) -
        \cos(\varphi_{e^+} - \varphi_{e^-}))\right)}, \\
    \varphi_{1} =& \pi \hspace{0.5cm} \varphi_{e^+} = \pi + 0.2 \hspace{0.5cm}
    \varphi_{e^-} = -(\pi + 0.2), \\
    y_1=\Delta& \hspace{0.5cm} y_2=-\Delta \hspace{0.5cm} y_{e^+} = \Delta
    \hspace{0.5cm} y_{e^-} = \Delta - 1.5.
  \end{split}
\end{align}
The matrix element squared divided by $\hat s^2$ tends to a constant when the
rapidity separation of the two outgoing partons grows large.  This is as
expected from BFKL and Regge theory.  Fig.~\ref{fig:twojets} also shows the separate contributions to the
total matrix element squared coming from the $\zg$ emission from the forward
moving quark line (black, dashed) and emission from the backward moving quark
line (green, dotted).  In this phase space slice, the leptons also have an
increasing positive rapidity and so the forward emission matrix element
describes the full matrix element most closely, with the contribution from
backward-emission falling at large values of $\Delta y$.  The sum of the forward
and backward emission matrix elements neglecting interference (magenta, dotted)
significantly overestimates the final result.  Once the (destructive)
interference effects have been taken into account, the full sum (red, solid)
correctly reproduces the LO matrix element (blue, thick solid).  It is therefore
clear that at low rapidities the inclusion of the interference effect plays an
important role in the accuracy of the matrix element.  Neither this effect nor
the interference between the $Z$ and $\gamma^*$ channels is included when
electroweak corrections are included in a parton
shower~\cite{Christiansen:2014kba,Krauss:2014yaa,Christiansen:2015jpa}.

\begin{figure}[!bt]
          \centering
          \begin{subfigure}[b]{0.48\textwidth}
                  \includegraphics[width=\textwidth]{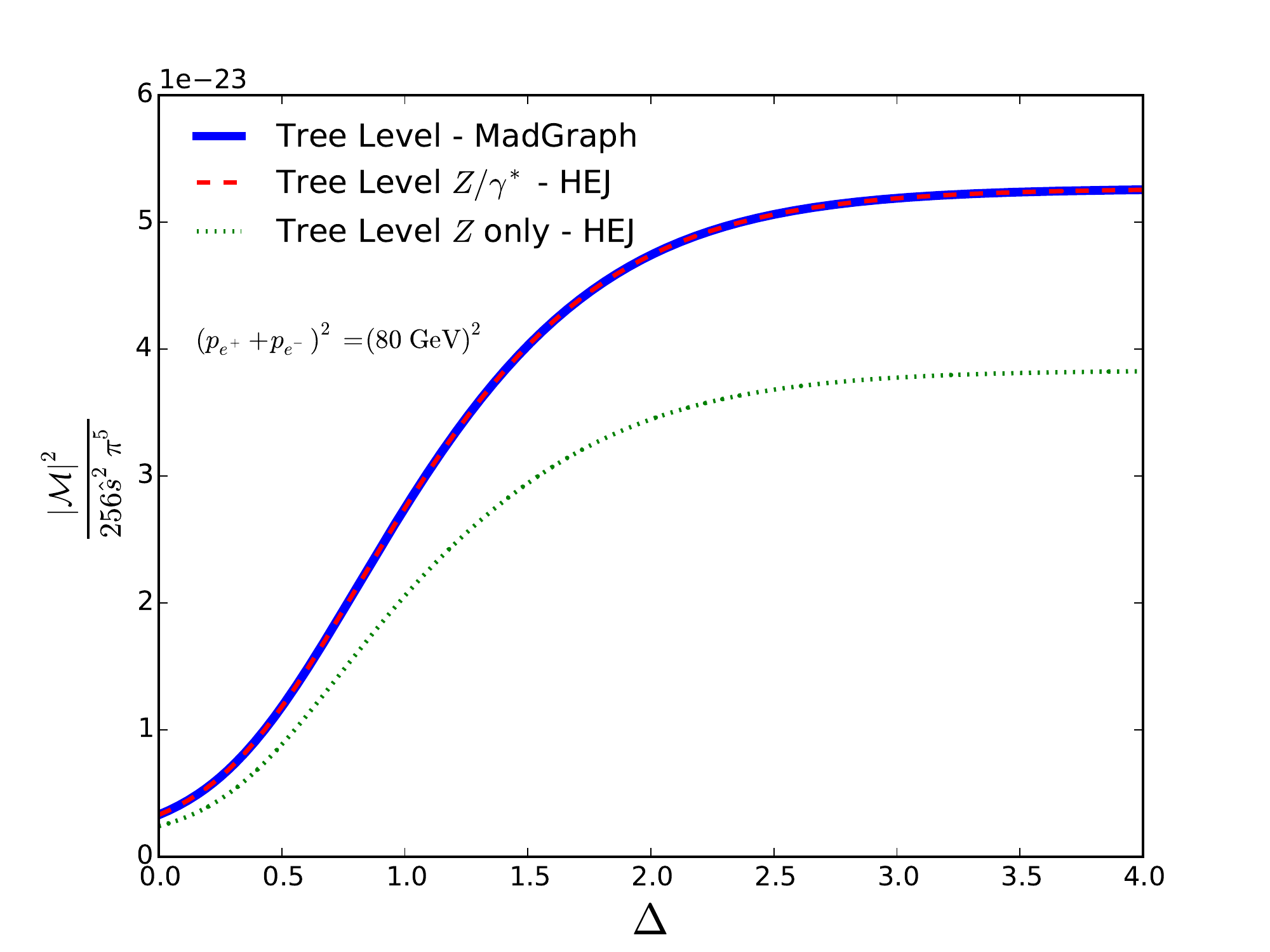}
                  \caption{}
                  \label{fig:LowDileptonMass}
          \end{subfigure}
          ~
          \begin{subfigure}[b]{0.48\textwidth}
                  \includegraphics[width=\textwidth]{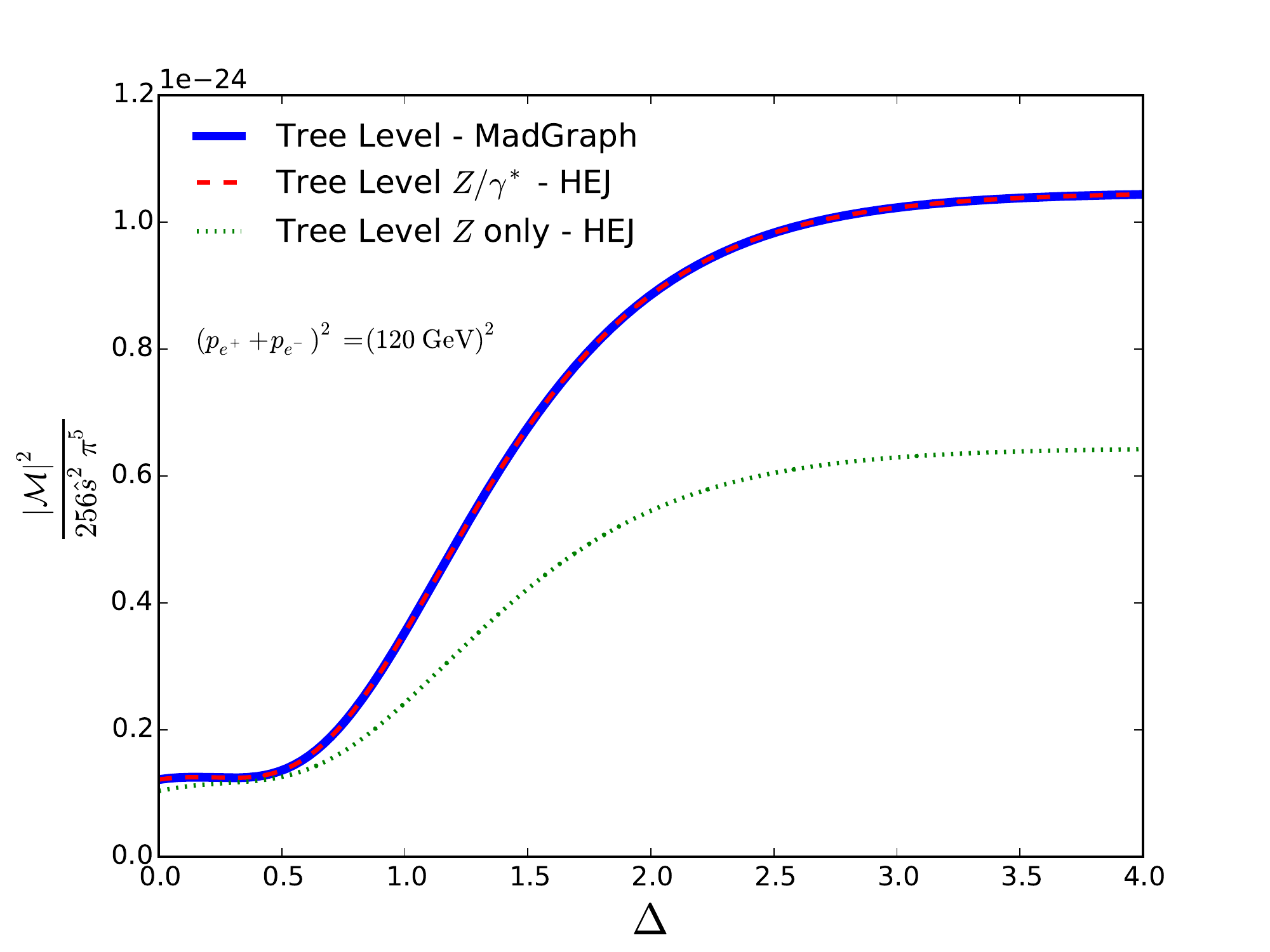}
                  \caption{}
                  \label{fig:HighDileptonMass}
          \end{subfigure}
          \caption{The matrix-element squared divided by the square of the
            partonic centre-of-mass energy for $qQ\to \zg qQ$ with the $\zg$ decaying
            to an electron-positron pair.  The
            $\mathcal{O}(\alpha_s^2 \alpha_W)$ tree-level contribution
            as described in HEJ (red, dashed) exactly matches that of
            Madgraph (blue, solid).  The terms corresponding to the production
            of a $Z$ boson only (green, dotted) significantly undershoots the
            full result.  The
            virtual photon terms are, therefore, clearly an important
            contribution to the matrix element away from the $Z$ Breit-Wigner
            peak.}
          \label{fig:DileptonMass}
  \end{figure}

One can also investigate the importance of the virtual photon contributions we
include and their interference with the pure $Z$ process.  The inclusion of the
virtual photon terms is particularly important when studying a combined lepton
invariant mass, $(p_{e^+} + p_{e^-})^2$, far from the $Z$ mass peak.  This can
be seen in Fig.~\ref{fig:DileptonMass}, where slices through phase space are
shown similarly to Fig.~\ref{fig:twojets}, but now for an (a) lower and (b) higher
value of the dilepton mass.  In both cases, the contribution of the virtual
photon processes is above 25\%.

Having established our description of the $2\to \zg +2$ parton process, we now turn
our attention to adding the all-order real corrections.  Our all-order
expression will take the form of a sum of terms like eq.~(\ref{eq:normal}) for
each of the three terms in eq.~(\ref{eqn:interference}), such that the squared
matrix element for $qQ\to (\zg\to ) e^+e^- q (n-2)g Q$ is:
\begin{align}
  \label{eq:allorderreal}
  \begin{split}
    |\mathcal{M}^{HE}_{qQ\to \zg q(n-2)gQ}|^2 =\ g_s^2& \frac{C_F}{8N_c}\ ( g_s^2
    \Ca)^{n-2}\  \\  \times \Bigg(& \frac{| j_a^{\zg}\cdot j_b|^2}{t_{a1}t_{a(n-1)}} \prod^{n-2}_{i=1} \frac{-V^2(q_{ai},
      q_{a(i+1)})}{t_{ai} t_{a(i+1)}}+\ \frac{|j_a \cdot j_b^{\zg} |^2}{t_{b1}t_{b(n-1)}}
    \prod^{n-2}_{i=1}\frac{-V^2(q_{bi}, q_{b(i+1)})}{t_{bi} t_{b(i+1)}}  \\
    &- \frac{2\Re\{(j_a^{\zg}\cdot j_b)(\overline{j_a \cdot
        j_b^{\zg}})\}}{\sqrt{t_{a1}t_{b1}}\sqrt{t_{a(n-1)} t_{b(n-1)}}}
    \prod^{n-2}_{i=1}\frac{V(q_{ai}, q_{a(i+1)})\cdot V(q_{bi},
      q_{b(i+1)})}{\sqrt{t_{ai}t_{bi}} \sqrt{t_{a(i+1)}t_{b(i+1)}}}\Bigg).
  \end{split}
\end{align}
In the case of $n=2$, this reduces back to eq.~(\ref{eqn:interference}).  If
either $a$ or $b$ is an incoming gluon, there is once again a unique set of
$t$-channel momenta and one can set the relevant $j_a^{\zg}$ or $j_b^{\zg}$ to
zero in the formula above.  This then gives eq.~(\ref{eq:normal}) up to a factor
of $\Ca/C_F$ which corrects the colour factor.

We therefore have a compact expression for the real-emission contribution to a
given process at any order in $\alpha_s$. All real corrections can then be
added by summing over $n\ge 2$, provided that each contribution is finite.  We
will organise the cancellation of singularities using a phase-space slicing
method which we describe in the next section.


\section{Virtual Corrections and the Cancellation of Divergences}
\label{sec:Constructing}

In the previous section, we derived a description for the dominant real emission
corrections in the HE limit for a given process contributing to $\zg$ plus
jets.  Here we describe the corresponding virtual corrections and the
organisation of the cancellation of divergences.

For a general QCD amplitude, the \emph{Lipatov Ansatz} gives an elegant
prescription for the leading logarithmic and next-to-leading logarithmic terms
of the virtual corrections in the HE limit~\cite{Kuraev:1976ge}.  Each
$t$-channel pole is supplemented with the following exponential factor:
\begin{align}
  \label{eq:lipatovansatz}
  \frac1{t_i} \longrightarrow \frac1{t_i} \exp( \hat{\alpha}(q_{i\perp}) (y_{i+1}-y_i)),
  \qquad \hat{\alpha}(q_{i\perp}) = -g_s^2 \Ca
  \frac{\Gamma(1-\varepsilon)}{(4\pi)^{2+\varepsilon}} \frac2\varepsilon \left(
  \frac{q_{i\perp}^2}{\mu^2} \right) ^\varepsilon,
\end{align}
where $q_{i\perp}$ is the transverse components of the relevant $t$-channel
momentum and we have used dimensional regularisation with $d=4+2\varepsilon$.
Given the different `$t$'s which enter the different terms of
eq.~(\ref{eq:allorderreal}), it is clear we must now also calculate the virtual
corrections in three separate terms.  We define $\Delta y_i=y_{i+1}-y_i$ and
then incorporate the all-order virtual corrections as follows:
\begin{align}
  \label{eq:allorderall}
  \begin{split}
    |\mathcal{M}^{HEJ}_{qQ\to \zg q(n-2)gQ}|^2 &=\ g_s^2 \frac{C_F}{8N_c}\ ( g_s^2
    \Ca)^{n-2}\  \\  \times \Bigg(& \frac{| j_a^\zg\cdot
      j_b|^2}{t_{a1}t_{a(n-1)}}
    \exp(2\hat{\alpha}(q_{a(n-1)\perp})\Delta y_{n-1}) \prod^{n-2}_{i=1} \frac{-V^2(q_{ai},
      q_{a(i+1)})}{t_{ai} t_{a(i+1)}} \exp(2\hat{\alpha}(q_{ai\perp})\Delta y_i)\\
    +\ &\frac{|j_a \cdot j_b^\zg |^2}{t_{b1}t_{b(n-1)}} \exp(2\hat{\alpha}(q_{b(n-1)\perp})\Delta y_{n-1})
    \prod^{n-2}_{i=1}\frac{-V^2(q_{bi}, q_{b(i+1)})}{t_{bi} t_{b(i+1)}} \exp(2\hat{\alpha}(q_{bi\perp})\Delta y_i) \\
    -\ &\frac{2\Re\{ (j_a^\zg\cdot j_b)(\overline{j_a \cdot
        j_b^\zg})\}}{\sqrt{t_{a1}t_{b1}}\sqrt{t_{a(n-1)} t_{b(n-1)}}} \exp((\hat{\alpha}(q_{a(n-1)\perp})+\hat{\alpha}(q_{b(n-1)\perp}))\Delta y_{n-1})\\
    & \; \prod^{n-2}_{i=1}\frac{V(q_{ai}, q_{a(i+1)})\cdot V(q_{bi},
      q_{b(i+1)})}{\sqrt{t_{ai}t_{bi}} \sqrt{t_{a(i+1)}t_{b(i+1)}}}\exp((\hat{\alpha}(q_{ai\perp})+\hat{\alpha}(q_{bi\perp}))\Delta y_{i})\Bigg).
  \end{split}
\end{align}
To find the physical result (cross section, distributions, etc.), we now need to
integrate over $n$-particle phase space and then sum over all $n\ge 2$.
However, before it is possible to do that, we must first organise the
cancellation of divergences.  There are two sources of divergences in
eq.~(\ref{eq:allorderall}): the poles in $\varepsilon$ within the virtual
corrections and, upon integration over all phase space, the divergences which
arise from any of the parton momenta going to zero.  We do not have collinear
singularities in our description, because by construction the particles are
assumed to be well-separated.

We will use a phase space slicing method in which we divide the available
phasespace into two regions by the introduction of a cut-off scale
$\lambda_{cut}$ on $p_\perp^2$.  Above the cut-off, we consider the emissions
`hard' and below the cut-off, we consider them to be `soft'.

The divergence arising from the emission of a soft gluon can be seen directly
from the effective vertex given in eq.~(\ref{eq:Veff}).  In the limit
$p_{i\perp}^2\to 0$, we find
\begin{align}
  \label{eq:Vlimit}
  -\frac{V^2(q_{i-1},q_{i})}{t_{i-1} t_i} \longrightarrow \frac{4}{p_{i\perp}^2},
  \qquad {\rm and} \qquad -\frac{V(q_{ai}, q_{a(i+1)})\cdot V(q_{bi},
      q_{b(i+1)})}{\sqrt{t_{ai}t_{bi}} \sqrt{t_{a(i+1)}t_{b(i+1)}}}
  \longrightarrow \frac{4}{p_{i\perp}^2}.
\end{align}
Therefore, the effect of the $i^{th}$ emitted parton becoming soft at the level
of the matrix element squared is:
\begin{equation}
  \lim_{p_i\rightarrow0} |\mathcal{M}^{HEJ}_{qQ\to \zg q(n-2)gQ}|^2\ =\
  \frac{4\Ca g_s^2}{|p_{i\perp}|^2}\ |\mathcal{M}^{HEJ}_{qQ\to \zg q(n-3)gQ}|^2,
\end{equation}
where the matrix element squared on the right-hand side is the corresponding one
for the momentum configuration of the matrix element on the left-hand side after
$p_i$ has been set to zero.  The relation is identical if either $q$ or $Q$ is replaced by a gluon.

The integration over the soft phase space for the
$i$th parton gives:
\begin{align}
  \label{eq:realint}
  \begin{split}
    \mu^{-2\epsilon} \int_{\rm
      soft}\frac{d^{3+2\epsilon}p_i}{(2\pi)^{3+2\epsilon}2E_i}\frac{4\Ca g_s^2}{|p_{i\perp}|^2} &=
    \mu^{-2\epsilon} \int_0^{\lambda_{cut}}\frac{d^{2+2\epsilon}p_{i\perp}}{(2\pi)^{2+2\epsilon}}
    \int_{y_{i-1}}^{y_{i+1}}\frac{dy_i}{4\pi} \frac{4\Ca g_s^2}{|p_{i\perp}|^2}\\
    &= \frac{4\Ca g_s^2\mu^{-2\epsilon}}{(2\pi)^{2+2\epsilon}4\pi}\
    (y_{i+1}-y_{i-1})\ \int_0^{\lambda_{cut}} \frac{d^{2+2\epsilon}p_{i\perp}}{|p_{i\perp}|^2}\\
    &= \frac{4\Ca g_s^2}{(2\pi)^{2+2\epsilon}4\pi}\ (y_{i+1}-y_{i-1})\
    \frac{1}{\epsilon} \frac{\pi^{1+\epsilon}}{\Gamma(\epsilon+1)} \left(\frac{\lambda_{cut}^2}{\mu^2}\right)^\epsilon
  \end{split}
\end{align}
where we have used a change of variables from $p_z$ to rapidity.  We will
eventually go on to integrate over the momenta of all other particles, but the
cancellation occurs already at the integrand level so we will not do so at this
point.  We have therefore found that the first-order correction to
the $qQ\to \zg q(n-3)gQ$ process from this soft
real emission is
\begin{align}
  \label{eq:realresult}
  \frac{\Ca g_s^2}{2^{2+2\epsilon}\pi^{2+\epsilon}}\ (y_{i+1}-y_{i-1})\
  \frac{1}{\epsilon\Gamma(1+\epsilon)}
  \left(\frac{\lambda_{cut}^2}{\mu^2}\right)^\epsilon \ \times\  |\mathcal{M}^{HEJ}_{qQ\to \zg q(n-3)gQ}|^2.
\end{align}
The corresponding first-order virtual correction is found by expanding the
exponentials in eq.~(\ref{eq:allorderall}).  We find
\begin{align}
  \label{eq:firstvirt}
  \begin{split}
   g_s^2 \frac{C_F}{8N_c}\ ( g_s^2
    \Ca)^{n-3} &\left(-g_s^2 \Ca\ \frac{\Gamma(1-\epsilon)}{2^{3+2\epsilon}
        \pi^{2+\epsilon}}\ \frac1{\epsilon}\ (y_{i+1}-y_{i-1}) \right) \\   \times \Bigg(& \frac{| j_a^\zg\cdot
      j_b|^2}{t_{a1}t_{a(n-1)}}     \left( \prod^{n-2}_{j=1, j\ne i} \frac{-V^2(q_{aj},
        q_{a(j+1)})}{t_{aj} t_{a(j+1)}} \right) \times
    2\left( \frac{q_{ai\perp}^2}{\mu^2} \right)^\epsilon  \\
    +\ &\frac{|j_a \cdot j_b^\zg |^2}{t_{b1}t_{b(n-1)}} \left(
    \prod^{n-2}_{j=1, j\ne i}\frac{-V^2(q_{bj}, q_{b(j+1)})}{t_{bj} t_{b(j+1)}}
  \right) \times  2\left( \frac{q_{bi\perp}^2}{\mu^2} \right)^\epsilon  \\
    -\ &\frac{2\Re\{ (j_a^\zg\cdot j_b)(\overline{j_a \cdot
        j_b^\zg})\}}{\sqrt{t_{a1}t_{b1}}\sqrt{t_{a(n-1)} t_{b(n-1)}}}
    \left(\prod^{n-2}_{j=1}\frac{V(q_{aj}, q_{a(j+1)})\cdot V(q_{bj},
      q_{b(j+1)})}{\sqrt{t_{aj}t_{bj}} \sqrt{t_{a(j+1)}t_{b(j+1)}}} \right) \\ &
  \times \left( \left( \frac{q_{ai\perp}^2}{\mu^2} \right)^\epsilon +  \left(
      \frac{q_{bi\perp}^2}{\mu^2} \right)^\epsilon \right)\Bigg).
  \end{split}
\end{align}

We can now go through term-by-term to show the divergences cancel and find the resulting
finite contribution to the matrix element squared.  For the backward line $\zg$ emission
squared terms, we have the following terms:
\begin{align}
  \begin{split}
    &g_s^2 \frac{C_F}{8N_c}\ ( g_s^2 \Ca)^{n-3} \frac{| j_a^\zg\cdot
      j_b|^2}{t_{a1}t_{a(n-1)}}     \left( \prod^{n-2}_{j=1, j\ne i} \frac{-V^2(q_{aj},
        q_{a(j+1)})}{t_{aj} t_{a(j+1)}} \right) \\
    &  \times \ \left(
      \frac{\Ca g_s^2}{2^{2+2\epsilon}\pi^{2+\epsilon}}\ (y_{i+1}-y_{i-1})\
      \frac{1}{\epsilon\Gamma(1+\epsilon)}
      \left(\frac{\lambda_{cut}^2}{\mu^2}\right)^\epsilon -  g_s^2 \Ca\
      \frac{\Gamma(1-\epsilon)}{2^{2+2\epsilon}\pi^{2+\epsilon}}\ \frac1{\epsilon}\
      (y_{i+1}-y_{i-1}) \left( \frac{q_{ai\perp}}{\mu} \right)^\epsilon \right) \\
    =\ & g_s^2 \frac{C_F}{8N_c}\ \frac{( g_s^2 \Ca)^{n-2}}{2^{2+2\epsilon}\pi^{2+\epsilon}}
    \frac{| j_a^\zg\cdot j_b|^2}{t_{a1}t_{a(n-1)}}     \left( \prod^{n-2}_{j=1, j\ne i} \frac{-V^2(q_{aj},
        q_{a(j+1)})}{t_{aj} t_{a(j+1)}} \right) (y_{i+1}-y_{i-1}) \\
    &  \times \ \left(
      \frac{1}{\epsilon\ \Gamma(1+\epsilon)}
      \left(\frac{\lambda_{cut}^2}{\mu^2}\right)^\epsilon - \
      \ \frac{\Gamma(1-\epsilon)}{\epsilon}\
      \left( \frac{q_{ai\perp}^2}{\mu^2} \right)^\epsilon \right).
  \end{split}
\end{align}
Performing the expansion in $\epsilon$ of the final bracket yields:
\begin{align}
  \label{eq:epsexpansion}
  \begin{split}
    &\Big((1 + \gamma_E\epsilon +
    \mathcal{O}(\epsilon^2))\Big(\frac{1}{\epsilon} +
    \ln\Big(\frac{\lambda_{cut}^2}{\mu^2}\Big) + \mathcal{O}(\epsilon)\Big) - (1
    + \gamma_E\epsilon + \mathcal{O}(\epsilon^2))\Big(\frac{1}{\epsilon} +
    \ln\Big(\frac{q_{ai\perp}^2}{\mu^2}\Big) + \mathcal{O}(\epsilon)\Big)\Big) \\
    &= \ln\left(\frac{\lambda_{cut}^2}{q_{ai\perp}^2}\right) +
    \mathcal{O}(\epsilon).
  \end{split}
\end{align}
The poles in $\epsilon$ and the $\gamma_E$ terms have identically cancelled
and we are left with a finite logarithm.  This is a similar form to that found
in~\cite{Andersen:2008gc,Andersen:2009nu}.  The procedure for the forward line
$\zg$ emission squared terms is identical and we find
\begin{align}
  g_s^2 \frac{C_F}{8N_c}\ \frac{( g_s^2 \Ca)^{n-2}}{2^{2+2\epsilon}\pi^{2+\epsilon}}
    \frac{| j_a\cdot j^\zg_b|^2}{t_{b1}t_{b(n-1)}}     \left( \prod^{n-2}_{j=1, j\ne i} \frac{-V^2(q_{bj},
        q_{b(j+1)})}{t_{bj} t_{b(j+1)}} \right) (y_{i+1}-y_{i-1}) \left( \ln\left(\frac{\lambda_{cut}^2}{q_{bi\perp}^2}\right) +
    \mathcal{O}(\epsilon) \right).
\end{align}
The cancellation for the interference terms is also similar and here we find
\begin{align}
  \label{eq:topsq}
  \begin{split}
    & -g_s^2 \frac{C_F}{8N_c}\ \frac{( g_s^2
      \Ca)^{n-2}}{2^{2+2\epsilon}\pi^{2+\epsilon}}\ \frac{2\Re\{ (j_a^\zg\cdot j_b)(\overline{j_a \cdot
        j_b^\zg})\}}{\sqrt{t_{a1}t_{b1}}\sqrt{t_{a(n-1)} t_{b(n-1)}}} \\
    & \qquad \times \left(\prod^{n-2}_{j=1}\frac{V(q_{aj}, q_{a(j+1)})\cdot V(q_{bj},
        q_{b(j+1)})}{\sqrt{t_{aj}t_{bj}} \sqrt{t_{a(j+1)}t_{b(j+1)}}} \right) \left(
      \ln\left(\frac{\lambda_{cut}^2}{\sqrt{q_{ai\perp}^2q_{bi\perp}^2}}\right) +
      \mathcal{O}(\epsilon) \right),
  \end{split}
\end{align}
as the finite remainder from the cancellation.  These results are valid for any
emission between the outer quarks/gluons which becomes soft.  If either of the
outer quarks/gluons becomes soft, this will also produce a divergence.  To
remain within the perturbative framework, we require that the outer particles
are constituents of the jets and that their transverse momentum is above a
minimum value.

It is clear that this result can be iterated order by order in $\alpha_s$.  We
would then form our final regulated all-order result as
\begin{align}
  \label{eq:allordereg}
  \begin{split}
    |\mathcal{M}^{HEJ-{\rm reg}}_{qQ\to \zg q(n-2)gQ}|^2 &=\ g_s^2 \frac{C_F}{8N_c}\ ( g_s^2
    \Ca)^{n-2}\  \\  \times \Bigg(& \frac{| j_a^{\zg}\cdot
      j_b|^2}{t_{a1}t_{a(n-1)}}
    \exp(\omega^0(q_{a(n-1)\perp})\Delta y_{n-1}) \prod^{n-2}_{i=1} \frac{-V^2(q_{ai},
      q_{a(i+1)})}{t_{ai} t_{a(i+1)}} \exp(\omega^0(q_{ai\perp})\Delta y_i)\\
    +\ &\frac{|j_a \cdot j_b^{\zg} |^2}{t_{b1}t_{b(n-1)}} \exp(\omega^0(q_{b(n-1)\perp})\Delta y_{n-1})
    \prod^{n-2}_{i=1}\frac{-V^2(q_{bi}, q_{b(i+1)})}{t_{bi} t_{b(i+1)}} \exp(\omega^0(q_{bi\perp})\Delta y_i) \\
    -\ &\frac{2\Re\{ (j_a^{\zg}\cdot j_b)(\overline{j_a \cdot
        j_b^{\zg}})\}}{\sqrt{t_{a1}t_{b1}}\sqrt{t_{a(n-1)} t_{b(n-1)}}} \exp(\omega^0(\sqrt{q_{a(n-1)\perp}q_{b(n-1)\perp}})\Delta y_{n-1})\\
    & \; \prod^{n-2}_{i=1}\frac{V(q_{ai}, q_{a(i+1)})\cdot V(q_{bi},
      q_{b(i+1)})}{\sqrt{t_{ai}t_{bi}} \sqrt{t_{a(i+1)}t_{b(i+1)}}}\exp(\omega^0(\sqrt{q_{ai\perp}q_{bi\perp}})\Delta y_{i})\Bigg),
  \end{split}
\end{align}
where we have defined
\begin{align}
  \label{eq:omega0}
  \omega^0(q_{\perp}^2) = - \frac{g_s^2 \Ca}{4\pi^2} \log\left( \frac{q_\perp^2}{\lambda_{cut}^2}\right).
\end{align}
One can easily check by expansion that this correctly reproduces the results in 
eqs.~\eqref{eq:epsexpansion}--\eqref{eq:topsq}.  However, the limit we have used from
eq.~\eqref{eq:Vlimit} is a limit and not an exact identity.  We therefore have
to account for the difference between $-V^2(q_{i-1},q_{i})/(t_{i-1} t_i)$ and
its strict limit of $4/p_{i\perp}^2$ for values of $p_{i\perp}$ below
$\lambda_{cut}$.  In practice, we include this correction for
$c_{cut}<|p_\perp|<\lambda_{cut}$ with $c_{cut} = 0.2$~GeV and find stable
results around this value.  We demonstrate
that our numerical results are also insensitive to the precise value of $\lambda_{cut}$
in appendix~\ref{sec:indep-lambd}.

A total (differential) cross section can then be obtained by summing over all
values of $n$ and integrating over the full $n$-particle phase space, using an
efficient Monte Carlo sampling algorithm~\cite{Andersen:2008ue,Andersen:2008gc}:
\begin{align}
  \label{eq:sigma}
  \begin{split}
    \sigma =& \sum_{f_a,f_b} \sum_{n=2}^\infty  \left( \prod_{i=1}^n \int \frac{d^3 p_i}{(2\pi)^3
        2E_i} \right) \int \frac{d^3 p_{e^-}}{(2\pi)^3 2E_{e^-}} \int \frac{d^3 p_{e^+}}{(2\pi)^3 2E_{e^+}} \\
    & \ \times (2\pi)^4 \delta^{(2)}\left(\sum_i p_{i\perp} - p_{e^-\perp} -p_{e^+\perp}\right) \\
    & \ \times \ |\mathcal{M}^{HEJ-{\rm reg}}_{f_af_b\to \zg
      f_a(n-2)gf_b}(\{p_i\},p_{e^-},p_{e^+})|^2 \ \frac{ x_a f_{f_a}(x_a, Q_a) x_b
      f_{f_b}(x_b,Q_b)}{\hat s^2}\ \Theta_{\rm cut},
  \end{split}
\end{align}
where $x_{a,b}$ are the momentum fractions of the incoming partons and
$f_{f_k}(x_k,Q_k)$ are the corresponding parton density functions for beam
$(k)$ and flavour $f_k$.  The factor of $\hat s^2$ is the usual phase space
factor.  The function $\Theta_{\rm cut}$ imposes any desired cuts on the final
state.  The minimum requirement is that the final state momenta cluster into at
least two jets for the desired algorithm\footnote{We use
  FastJet~\cite{Cacciari:2011ma} within our code and so are compatible with
  (almost) any choice of jet algorithm and parameter.}.

In the regions of phase space where all final state particles are well separated
in rapidity, this gives the dominant terms in QCD at all orders in $\alpha_s$  (the leading logarithmic
terms in $s/t$).  However, in other areas of
phase space, the differences due to the approximations used in
$|\mathcal{M}^{HEJ-{\rm reg}}_{qQ\to \zg q(n-2)gQ}|^2$ will become more
significant.  We can therefore further improve upon eq.~\eqref{eq:sigma} by
matching our results to fixed order results.  Here, we match to
high-multiplicity tree-level
results obtained from Madgraph5\_aMC@NLO~\cite{Alwall:2014hca} in two different ways.
This amounts to merging tree-level samples of different orders according to the
logarithmic prescription of HEJ.
\begin{enumerate}
\item Matching for FKL configurations

  As described in section~\ref{sec:HEQCD}, these are the particle assignments
  and momentum configurations which contain the dominant leading-logarithmic
  terms in $s/t$.  The first step of the HEJ description was to develop an
  approximation to the matrix element for these processes which was later
  supplemented with the finite correction which remained after cancelling the
  real and virtual divergences: $\overline{|\mathcal{M}_{qg\to Zqg}^{HE}|}^2$
  (eq.~(\ref{eq:qgamp})) or $\overline{|\mathcal{M}_{qQ\to ZqQ}^{HE}|}^2$
  (eq.~(\ref{eq:allorderreal})).  The approximation is necessary to allow us to
  describe the matrix element for any (and in particular, large) $n$ and for
  including both the leading real and virtual corrections.  However,
  if the parton momenta cluster into four or fewer \emph{jets}\footnote{These
    may have arisen from many more partons.}, the full tree-level matrix element remains
  calculable.  In these cases, we perform the matching multiplicatively, so we
  multiply the integrand of eq.~(\ref{eq:sigma}) by
  \begin{align}
    |\mathcal{M}^{\rm full}_{qQ\to \zg q(k-2)gQ}(p_a,p_b,\{
    j_i'\})|^2/|\mathcal{M}^{HEJ}_{qQ\to \zg q(k-2)gQ}(p_a,p_b,\{j_i'\})|^2.
  \end{align}
  Here, $\{j_i'\}$ are the jet momenta after a small amount of
  reshuffling.  This is necessary because the evaluation of the tree-level matrix elements
  assumes that the jet momenta are both on-shell and have transverse momenta which
  sum to zero, neither of which is true in general for our events due to the
  presence of extra emissions.  Our reshuffling algorithm~\cite{Andersen:2011hs} redistributes this
  extra transverse momentum in proportion to the size of the transverse
  momentum of each jet.  The plus and minus light-cone components are then adjusted such
  that the jet is put on-shell and the rapidity remains unaltered.  This last
  feature ensures that after reshuffling the event is still in an FKL
  configuration. 

  After this multiplicative matching factor has been included, the regularisation then proceeds as before.

\item Matching for non-FKL configurations

  Away from regions in phase space where the quarks and gluons are
  well-separated, the non-FKL configurations will play a more significant
  r\^ole.  These have so far not been accounted for at all, and hence we add
  three exclusive samples of leading-order two-jet, three-jet and four-jet
  leading-order events to our resummed events.  The distinction between the
  samples is made following the choice of jet algorithm and parameters.

\end{enumerate}

These two matching schemes complete our description of the production of $\zg$
with at least two jets, including the leading high-energy logarithms at all
orders in $\alpha_s$.  In the next two sections, we compare the predictions from
this formalism to LHC data.


\section{Comparisons to LHC Data}
\label{sec:Comparisons}

\subsection{ATLAS - $Z$+Jets Measurements}
\label{sub:ATLAS}

We now compare the results of the formalism described in the previous sections
to data.  We begin with a recent ATLAS analysis of $Z$-plus-jets events from
7~TeV collisions~\cite{Aad:2013ysa}.  We summarise the cuts in table~\ref{tab:atlascuts}.
\begin{table}[hbt]
  \centering
  \begin{tabular}{|l|c|}
    \hline
    Lepton Cuts & $p_{T\ell}>20$~GeV, \; $|\eta_\ell|<2.5$ \\
    & $\Delta R^{\ell^+\ell^-} > 0.2$, \; $66$~GeV $\leq m^{\ell^+\ell^-} \leq
      116$~GeV \\ \hline
    Jet Cuts (anti-$k_T$, 0.4) & $p_{Tj}>30$~GeV, \; $|y_j|<4.4$ \\
    & $\Delta R^{j\ell} >0.5$ \\
\hline
  \end{tabular}
  \caption{The cuts applied to the theory simulations in the ATLAS
    $Z$-plus-jets analysis results shown in Figs.~\ref{fig:ATLAS_2a}--\ref{fig:ATLAS_7b}.}
  \label{tab:atlascuts}
\end{table}
Any jet which failed the jet-lepton isolation cut was removed from the event, but the
event itself is kept provided there are a sufficient number of other jets
present.  Throughout, the central value of the HEJ predictions has been
calculated with factorisation and renormalisation scales set to
$\mu_F=\mu_R=H_T/2$, and the theoretical uncertainty band has been determined by
varying these independently by up to a factor of 2 in each direction (removing
the corners where the relative ratio is greater than two).  Also shown in the
plots taken from the ATLAS paper are theory predictions from
Alpgen~\cite{Mangano:2002ea}, Sherpa~\cite{Gleisberg:2008ta,Hoeche:2012yf},
MC@NLO~\cite{Frixione:2002ik} and
BlackHat+Sherpa~\cite{Berger:2010vm,Ita:2011wn}.   We will also comment on the
recent theory description of Ref.\cite{Frederix:2015eii}.

\begin{figure}[bt]
  \centering
  \begin{subfigure}[b]{0.48\textwidth}
    \includegraphics[width=0.92\textwidth]{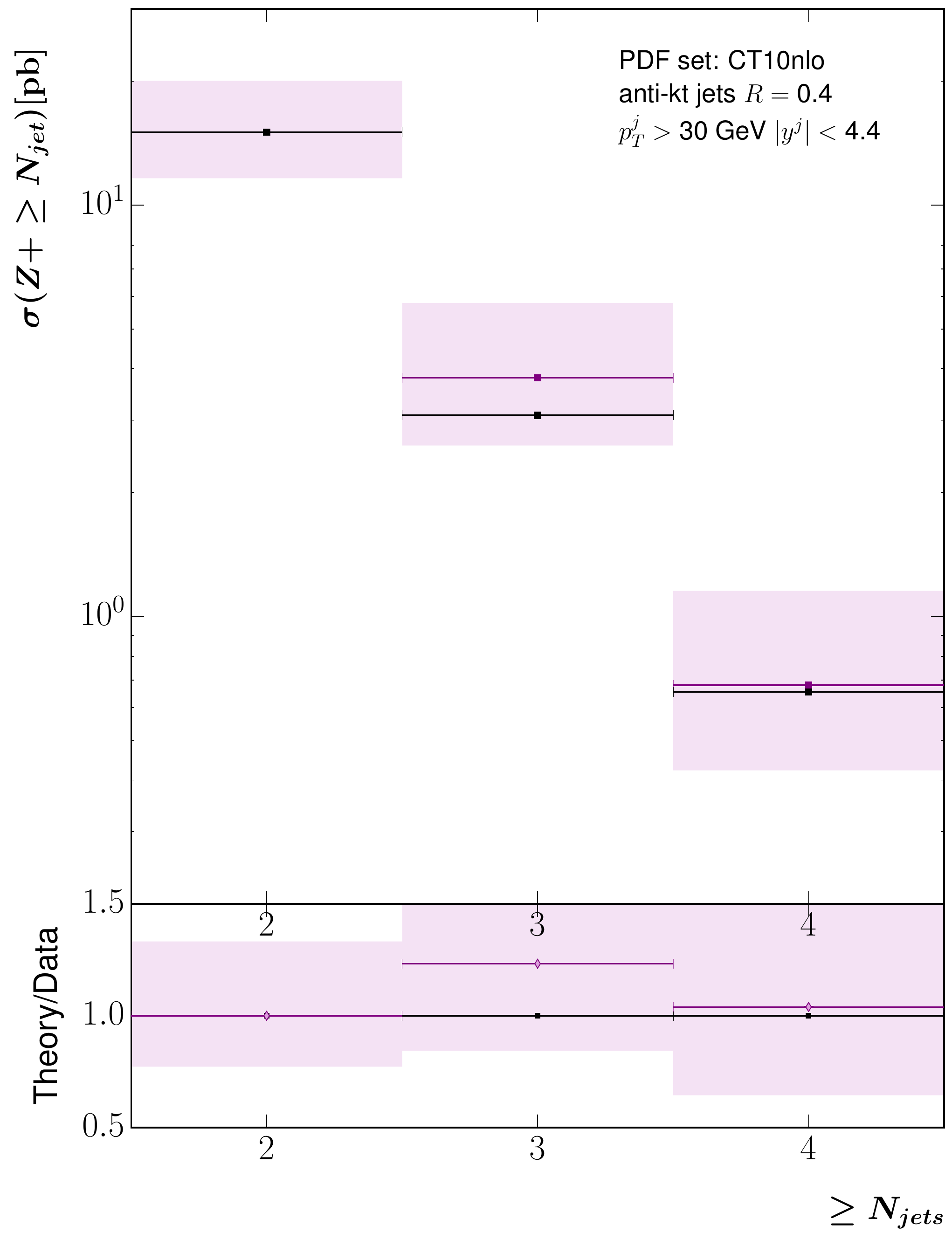}
    \caption{}
    \label{fig:HEJ_ATLAS_2a}
  \end{subfigure}
  ~
  \begin{subfigure}[b]{0.48\textwidth}
    \includegraphics[width=0.9\textwidth]{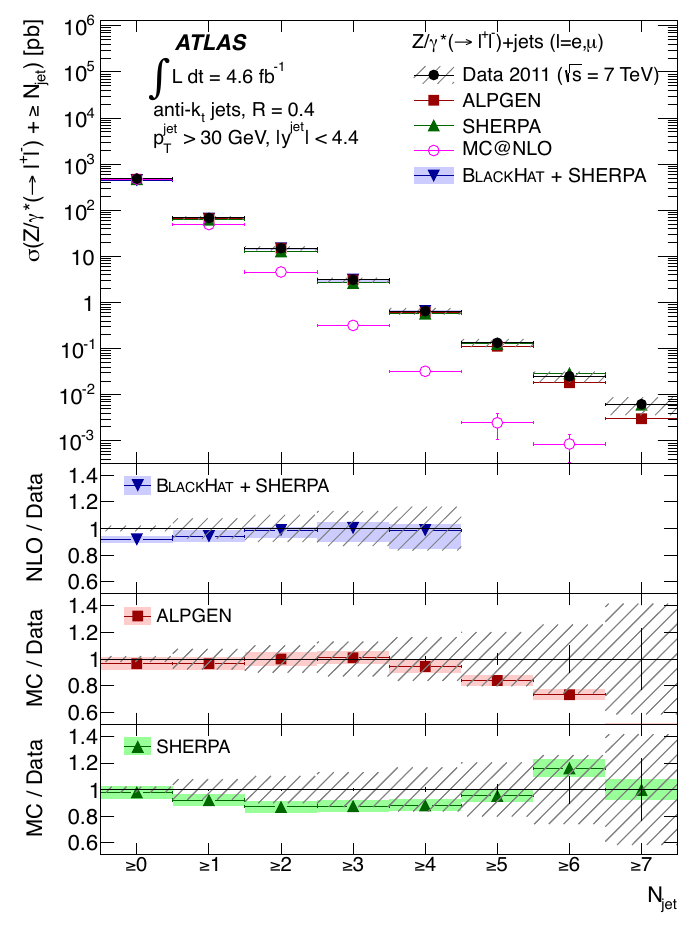}
    \caption{}
    \label{fig:MC_ATLAS_2a}
  \end{subfigure}
  \caption{These plots show the inclusive jet rates from (a) HEJ and (b) other
    theory descriptions and data~\cite{Aad:2013ysa}.  HEJ events all contain at
    least two jets and do not contain matching for 5 jets and above, so these
    bins are not shown.}
  \label{fig:ATLAS_2a}
\end{figure}

In Fig.~\ref{fig:ATLAS_2a} we begin this set of comparisons with predictions
and measurements of the inclusive jet rates.  HEJ and most of the other theory
frameworks give a reasonable description of these rates.  The MC@NLO
prediction drops below the data because it only contains the hard-scattering
matrix element for $\zg$ production and relies on a parton shower for additional
emissions beyond the one hard jet. The HEJ predictions have a larger uncertainty band which largely
arises from the use of leading-order results in the matching procedures.

We will now discuss a number of the differential distributions.  In
Ref.~\cite{Aad:2013ysa} these are displayed as distributions normalised to the
inclusive $\zg$-rate. However, given
the excellent agreement between the HEJ-prediction and data for the inclusive
2-jet cross section, we prefer to compare to data directly the prediction
obtained with HEJ for the distributions. 

The size of the scale variation of the HEJ predictions is largely dictated by
the matching to leading order accuracy. The smaller scale variation in the
results of e.g. Blackhat+Sherpa is therefore a reflection of the benefit of
going to NLO. The choice of not normalising the HEJ predictions further
increases the size of the scale variation bands, as there is no cancellation in
scale dependence in numerator and denominator.  We find, though, that our scale
dependence tends to lead to a change in overall normalisation rather than in
shape.  We demonstrate this by plotting
$(1/\sigma((\zg\to e^+e^-)+\ge 2j))\  d\sigma/dX$ for various variables $X$ in
appendix~\ref{sec:normalisation}.  Including such a normalisation factor
significantly reduces the size of the scale uncertainty band, down to less than
$\pm 10\%$ in both cases.  The quality of agreement with the central line is unchanged.

\begin{figure}[bt]
  \begin{subfigure}[b]{0.48\textwidth}
    \includegraphics[width=0.92\textwidth]{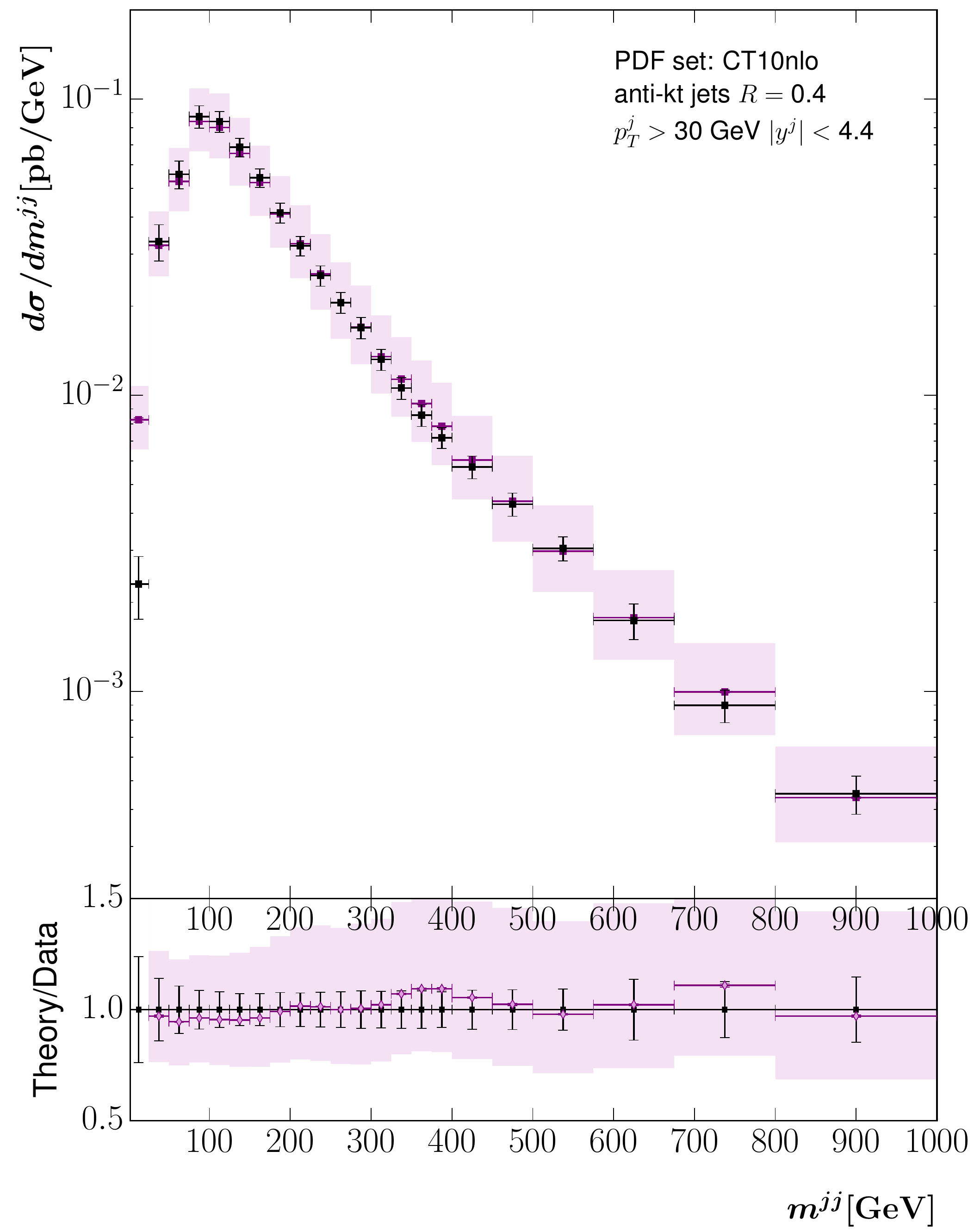}

    \vspace{0.2cm}
    \caption{}
    \label{fig:HEJ_ATLAS_11b}
  \end{subfigure}
  ~
  \begin{subfigure}[b]{0.48\textwidth}
    \includegraphics[width=0.9\textwidth]{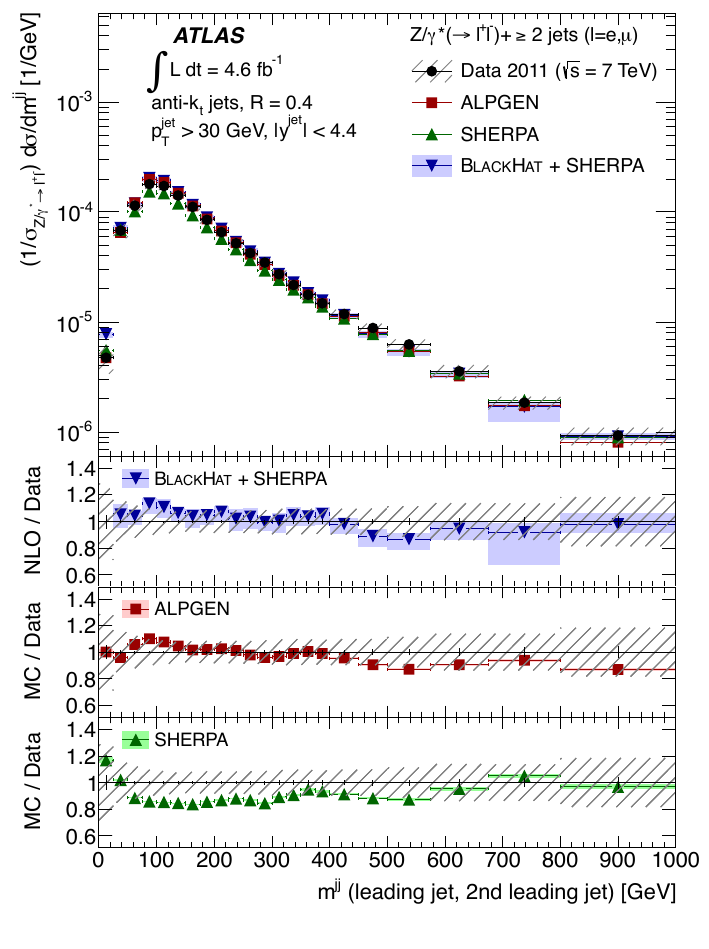}
    \caption{}
    \label{fig:MC_ATLAS_11b}
  \end{subfigure}
  \caption{These plots show the invariant mass between the leading and
    second-leading jet in $p_T$.  As in Fig.~\ref{fig:ATLAS_2a}, predictions are
    shown from (a) HEJ and (b) other theory descriptions and
    data~\cite{Aad:2013ysa}. These studies will inform Higgs plus dijets
    analyses, where cuts are usually applied to select events with large
    $m_{12}$.}
  \label{fig:ATLAS_11b}
\end{figure}

The first
differential distribution we consider here is the distribution of the invariant
mass between the two hardest jets, Fig.~\ref{fig:ATLAS_11b}.  The region of
large invariant mass is particularly important because this is a critical region
for studies of vector boson fusion (VBF) processes in Higgs-plus-dijets, and as
previously discussed, the corrections arising from QCD are similar in both
processes: The radiation patterns are largely universal between these processes,
so one can test the quality of theoretical descriptions in $\zg$-plus-dijets and
use these to inform the $Hjj$-analyses.  It is also a distribution which will be
studied to try to detect subtle signs of new physics.  In this study, HEJ and
the other approaches all give a good description of this variable out to
1~TeV. It will be interesting to see if the very good agreement between HEJ and
  the central data points will survive, once larger data sets lead to a
  reduction in the experimental uncertainty.  The merged sample of
Ref.~\cite{Frederix:2015eii} (Fig.~9 in that paper) combined with the Pythia8
parton shower performs reasonably well throughout the range with a few
deviations of more than 20\%, while that combined with Herwig++ deviates badly.
In a recent ATLAS analysis of $W$-plus-dijet events~\cite{Aad:2014qxa}, the
equivalent distribution was extended out to 2~TeV and almost all of the
theoretical predictions deviated significantly while the HEJ prediction remained
flat.  This is one region where the high-energy logarithms, included only in
HEJ, are expected to become large.

\begin{figure}[bt]
  \centering
  \begin{subfigure}[b]{0.48\textwidth}
    \includegraphics[width=0.91\textwidth]{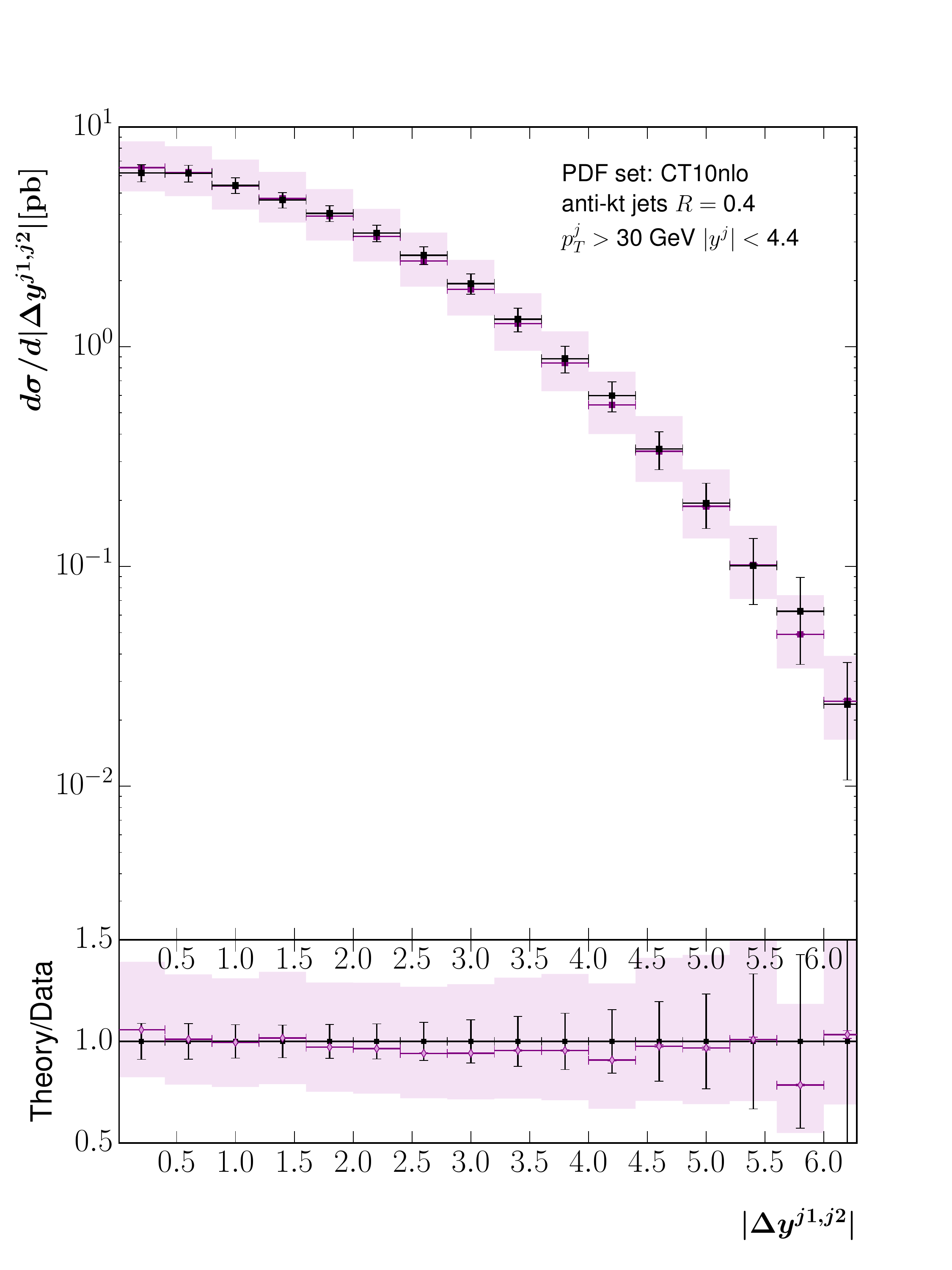}
    \caption{}
    \label{fig:HEJ_ATLAS_11a}
  \end{subfigure}
  ~
  \begin{subfigure}[b]{0.48\textwidth}
    \includegraphics[width=0.91\textwidth]{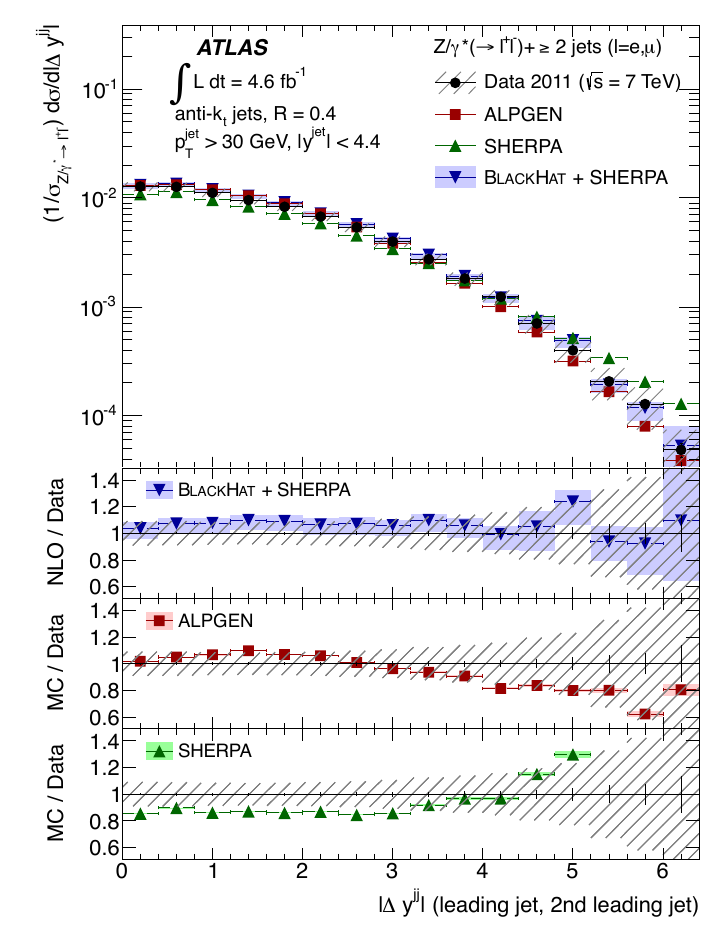}

    \vspace{0.1cm}
    \caption{}
    \label{fig:MC_ATLAS_11a}
  \end{subfigure}
  \caption{The comparison of (a) HEJ and (b) other theoretical descriptions and
    data~\cite{Aad:2013ysa} to
    the distribution of the absolute rapidity different between the two leading
    jets.  HEJ and Blackhat+Sherpa give the best description. These results
    will inform analyses of Higgs plus dijets, where cuts are usually applied
    to select events with large rapidity separation of jets.}
  \label{fig:ATLAS_11a}
\end{figure}

In Fig.~\ref{fig:ATLAS_11a}, we show the comparison of various theoretical
predictions to the distribution of the absolute rapidity difference between the
two leading jets.  It is clear in the left plot that HEJ gives an excellent
description of this distribution.  This is to some extent expected as
high-energy logarithms are associated with rapidity separations.  However, this
variable is only the rapidity separation between the two hardest jets which is
often not representative of the total rapidity `length' of events with more than
two hard jets, since the hardest jets tend to be central in rapidity.
Nonetheless, the HEJ description also performs well in this restricted scenario.  The
next-to-leading order (NLO) calculation of Blackhat+Sherpa also describes the
distribution quite well while the other merged, fixed-order samples deviate from
the data at larger values.  The merged samples of Ref.~\cite{Frederix:2015eii}
(Fig.~8 in that paper) describe this distribution well for small values of this
variable up to about 3 units when combined with Herwig++ and for most of the
range when combined with the Pythia8 parton shower, only deviating above 5 units.

\begin{figure}[bt]
  \begin{subfigure}[b]{0.48\textwidth}
    \includegraphics[width=0.94\textwidth]{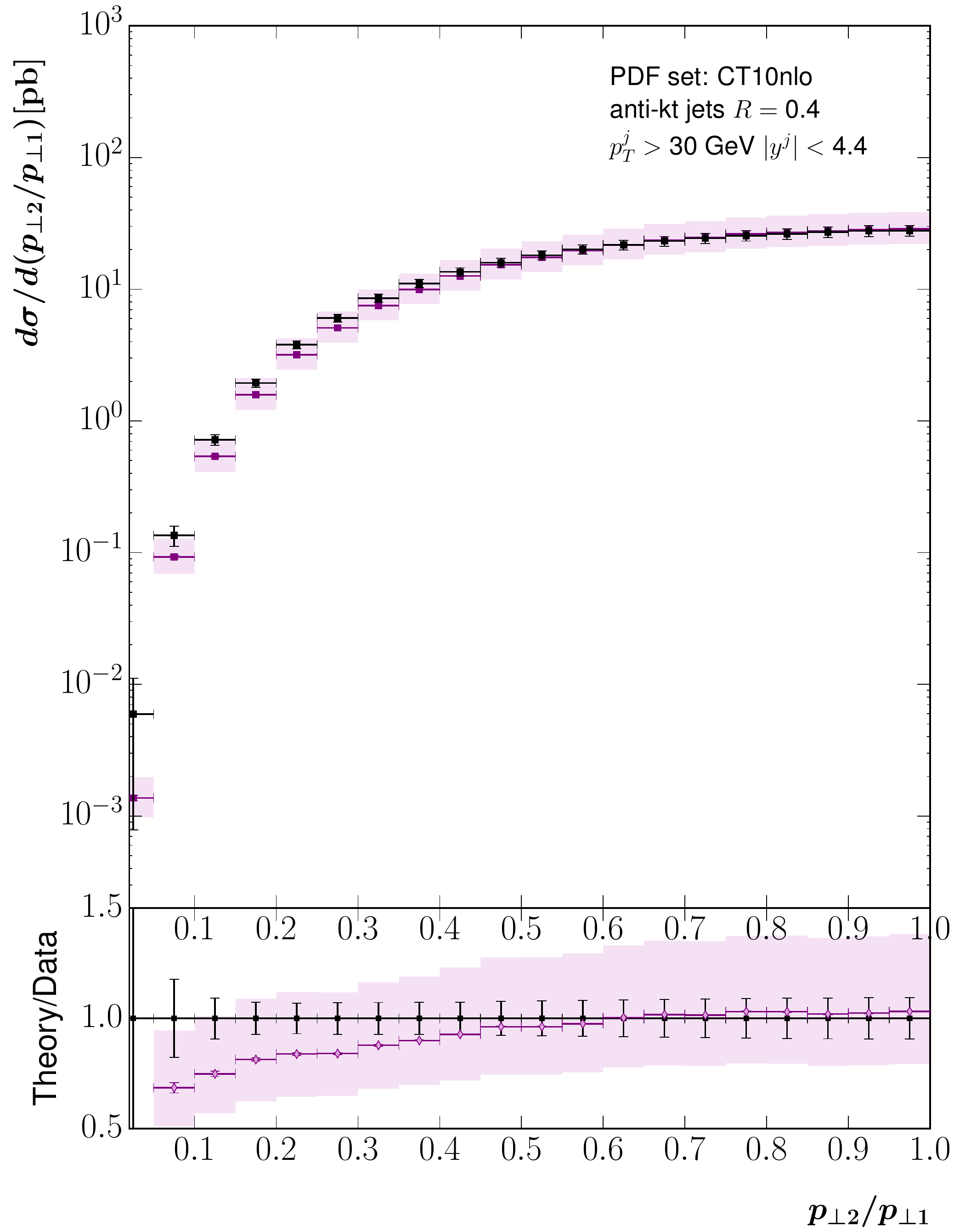}

    \vspace{0.1cm}
    \caption{}
    \label{fig:HEJ_ATLAS_7b}
  \end{subfigure}
  ~
  \begin{subfigure}[b]{0.48\textwidth}
    \includegraphics[width=0.9\textwidth]{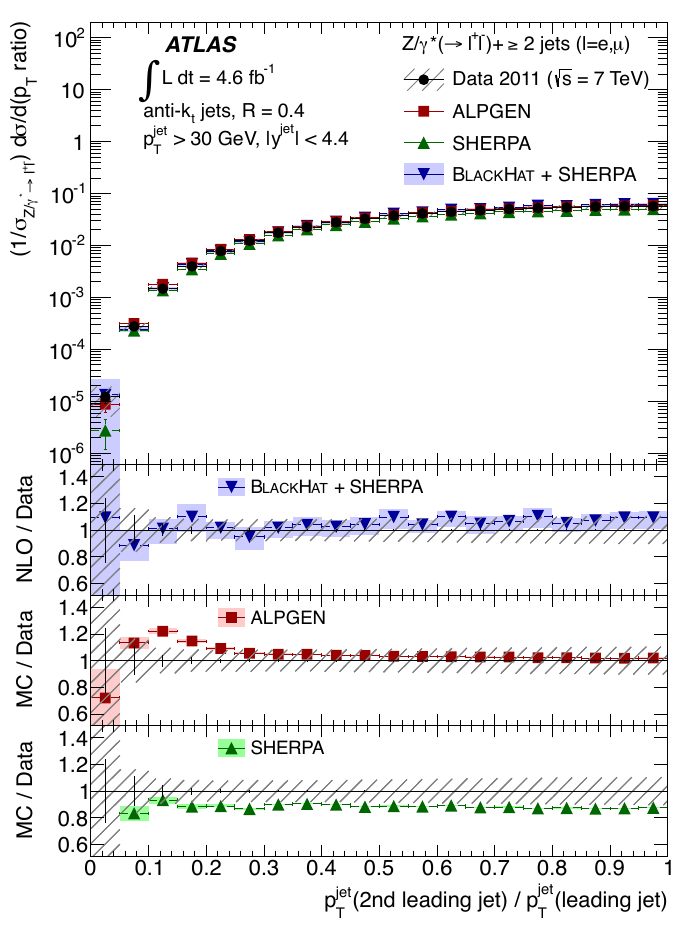}
    \caption{}
    \label{fig:MC_ATLAS_7b}
  \end{subfigure}
  \caption{These plots show the differential cross section in the ratio of the leading
     and second leading jet in $p_T$ from (a) HEJ and (b) other
    theory descriptions and data~\cite{Aad:2013ysa}.}
  \label{fig:ATLAS_7b}
\end{figure}

The final distribution in this section is that of the ratio of the transverse
momentum of the second hardest jet to the hardest jet.  The perturbative
description of HEJ does not contain any systematic evolution of transverse
momentum and this can be seen where its prediction undershoots the data at low
values of $p_{T2}/p_{T1}$.  However, for values of $p_{T2} \gtrsim 0.5 p_{T1}$,
the ratio of the HEJ prediction to data is extremely close to 1.  The
fixed-order based predictions shown in Fig.~\ref{fig:ATLAS_7b} are all fairly
flat above about 0.2, but the ratio to the data differs by about 10\% for the
Blackhat+Sherpa and Sherpa predictions.  Clearly the theoretical uncertainties
for the fixed-order based predictions for values of $p_{\perp 2}/p_{\perp 1}$
close to 1 are very small.  Comparing to the normalised distribution in
appendix~\ref{sec:normalisation}, this is a region where the theoretical
uncertainties in HEJ also become very small when normalisation is taken into
account.

\clearpage
\subsection{CMS - $Z$ + Jets Measurements}
\label{sub:CMS}

We now compare to data from a CMS analysis of events with a $\zg$ boson produced
in association with jets~\cite{Khachatryan:2014zya}.  We show, for comparison,
the plots from that analysis which contain theoretical predictions from
Sherpa~\cite{Gleisberg:2008ta,Hoeche:2012yf}, Powheg~\cite{Alioli:2010qp} and
MadGraph+Pythia~\cite{Alwall:2014hca}.  The cuts used for this analysis are summarised in
table~\ref{tab:cmscuts}.
\begin{table}[hbt]
  \centering
  \begin{tabular}{|l|c|}
    \hline
    Lepton Cuts & $p_{T\ell}>20$~GeV, \; $|\eta_\ell|<2.4$ \\
    &\; $71$~GeV $\leq m^{\ell^+\ell^-} \leq
      111$~GeV \\ \hline
    Jet Cuts (anti-$k_T$, 0.5) & $p_{Tj}>30$~GeV, \; $|y_j|<2.4$ \\
    & $\Delta R^{j\ell} >0.5$ \\
\hline
  \end{tabular}
  \caption{Cuts applied to theory simulations in the CMS
    $Z$-plus-jets analysis results shown in
    Figs.~\ref{fig:CMS_2a}--\ref{fig:CMS_3c}.}
  \label{tab:cmscuts}
\end{table}

As in the previous section, any jet which failed the final jet-lepton isolation cut was
removed from the event, but the event itself is kept provided there are a
sufficient number of other jets present.  The main difference to these cuts and
those of ATLAS in the previous section is that the jets are required to be more
central; $|y|<2.4$ as opposed to $|y|<4.4$.  This allows less room for
evolution in rapidity; however, as we will see, HEJ predictions are still relevant in this
scenario.  Once again, the central values are given by $\mu_F=\mu_R=H_T/2$ with
theoretical uncertainty bands determined by varying these independently by
factors of two around this value.  Once again, the theoretical
uncertainty bands on the HEJ predictions are large (we note that they are not
displayed in the MadGraph+Pythia6 predictions).  The size is dictated by matching to leading-order.  As illustrated in
appendix~\ref{sec:normalisation}, the scale variation effects are largely an overall normalisation
and not a change in shape and are significantly reduced in normalised
distributions.  Therefore the agreement between the central predictions and
data is more significant than the variation bands initially suggest.  HEJ events always contain a minimum of two
jets and therefore here we only compare to the distributions for an event sample
with at least two jets or above.

\begin{figure}[btp]
  \centering
  \begin{subfigure}[b]{0.46\textwidth}
    \includegraphics[width=0.84\textwidth]{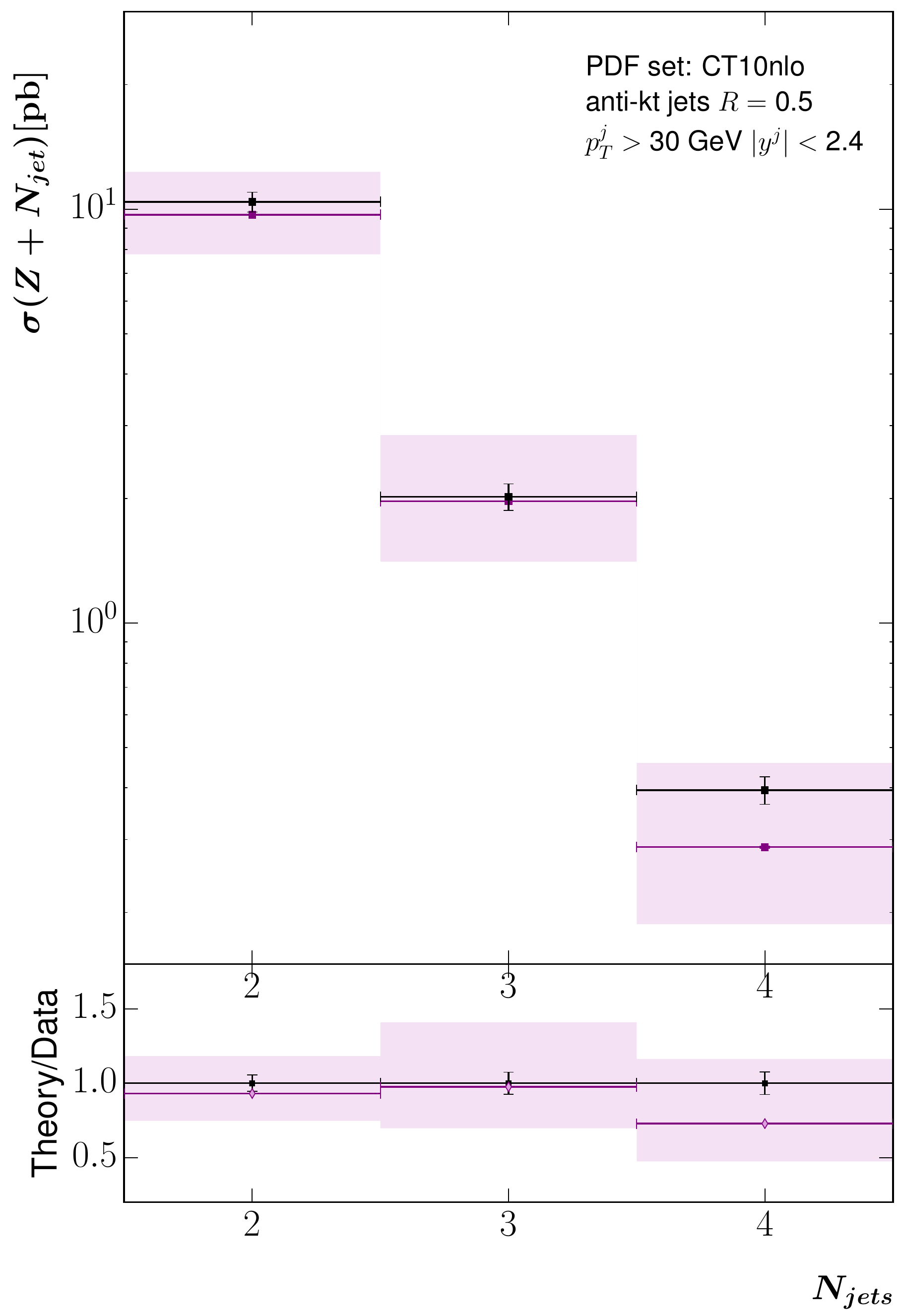}
    \caption{}
    \label{fig:HEJ_CMS_2a}
  \end{subfigure}
  ~
  \begin{subfigure}[b]{0.48\textwidth}
    \includegraphics[width=0.82\textwidth]{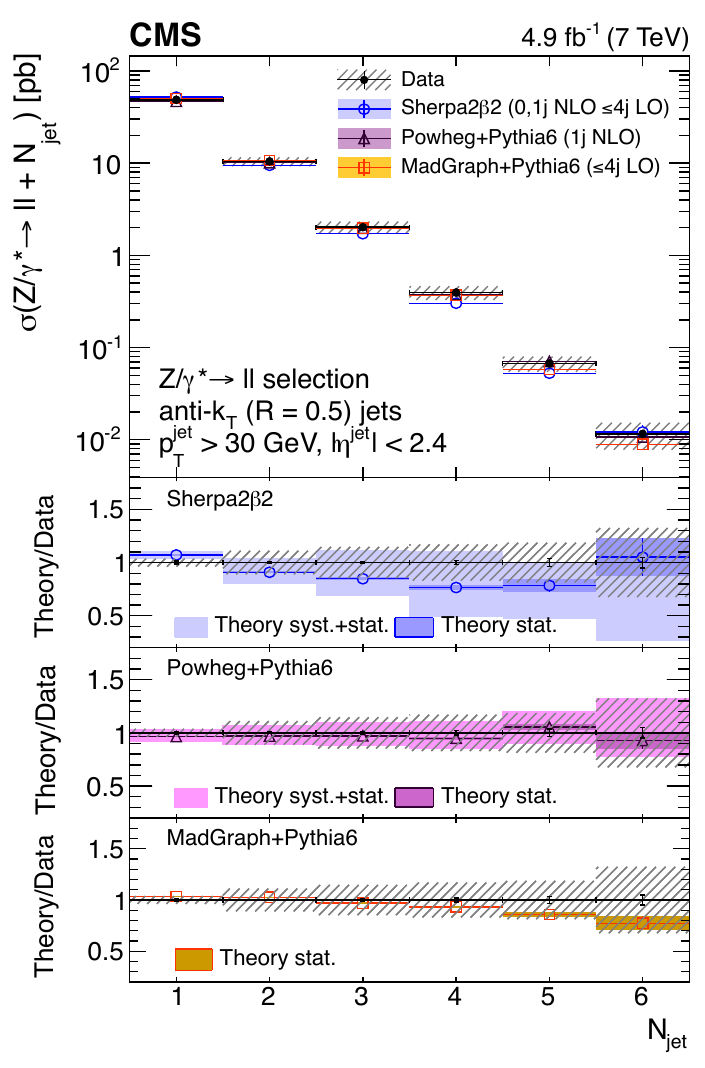}
    \caption{}
    \label{fig:MC_CMS_2a}
  \end{subfigure}
  \caption{The inclusive jet rates from~\cite{Khachatryan:2014zya} compared to
    predictions from (a) the HEJ description and (b) other theoretical
    descriptions.}
  \label{fig:CMS_2a}
\end{figure}

\begin{figure}
  \begin{subfigure}[b]{0.46\textwidth}
    \includegraphics[width=0.84\textwidth]{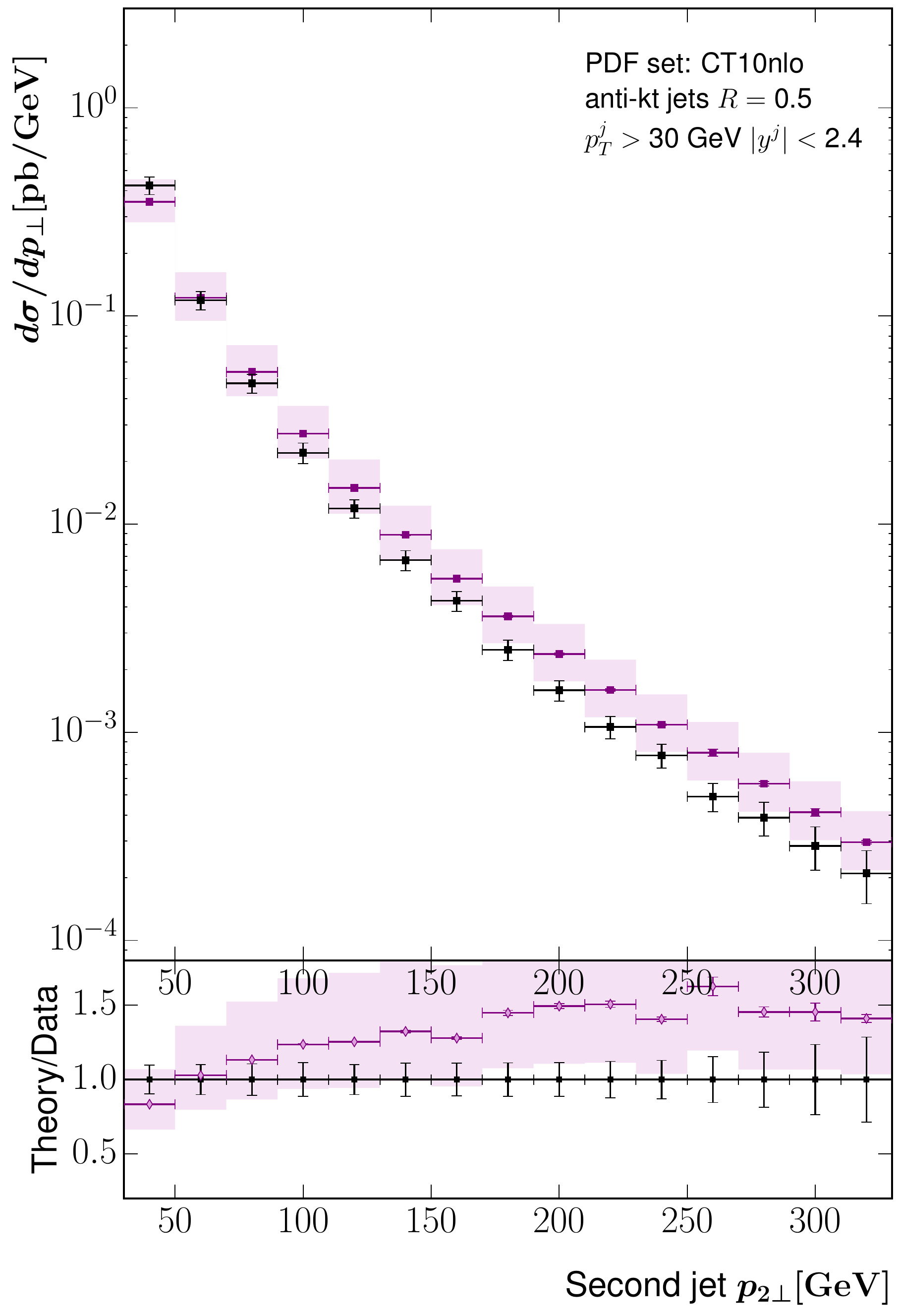}

    \vspace{0.2cm}
    \caption{}
    \label{fig:HEJ_CMS_7b-1}
  \end{subfigure}
  ~
  \begin{subfigure}[b]{0.48\textwidth}
    \includegraphics[width=0.82\textwidth]{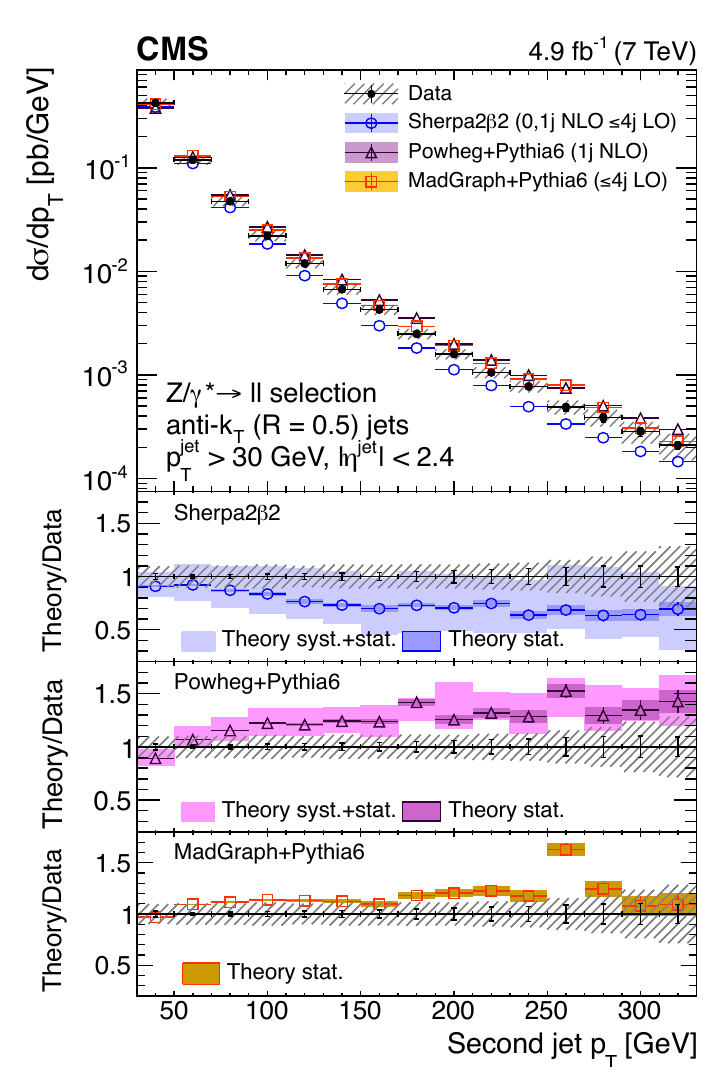}

    \vspace{0.2cm}
    \caption{}
    \label{fig:MC_CMS_7b}
  \end{subfigure}
  \caption{The transverse momentum distribution of the second hardest jet in
    inclusive dijet events in~\cite{Khachatryan:2014zya}, compared to (a) the
    predictions from HEJ and (b) the predictions from other theory descriptions.}
  \label{fig:CMS_3b}
\end{figure}

\begin{figure}[btp]
  \centering
  \begin{subfigure}[b]{0.46\textwidth}
    \includegraphics[width=0.86\textwidth]{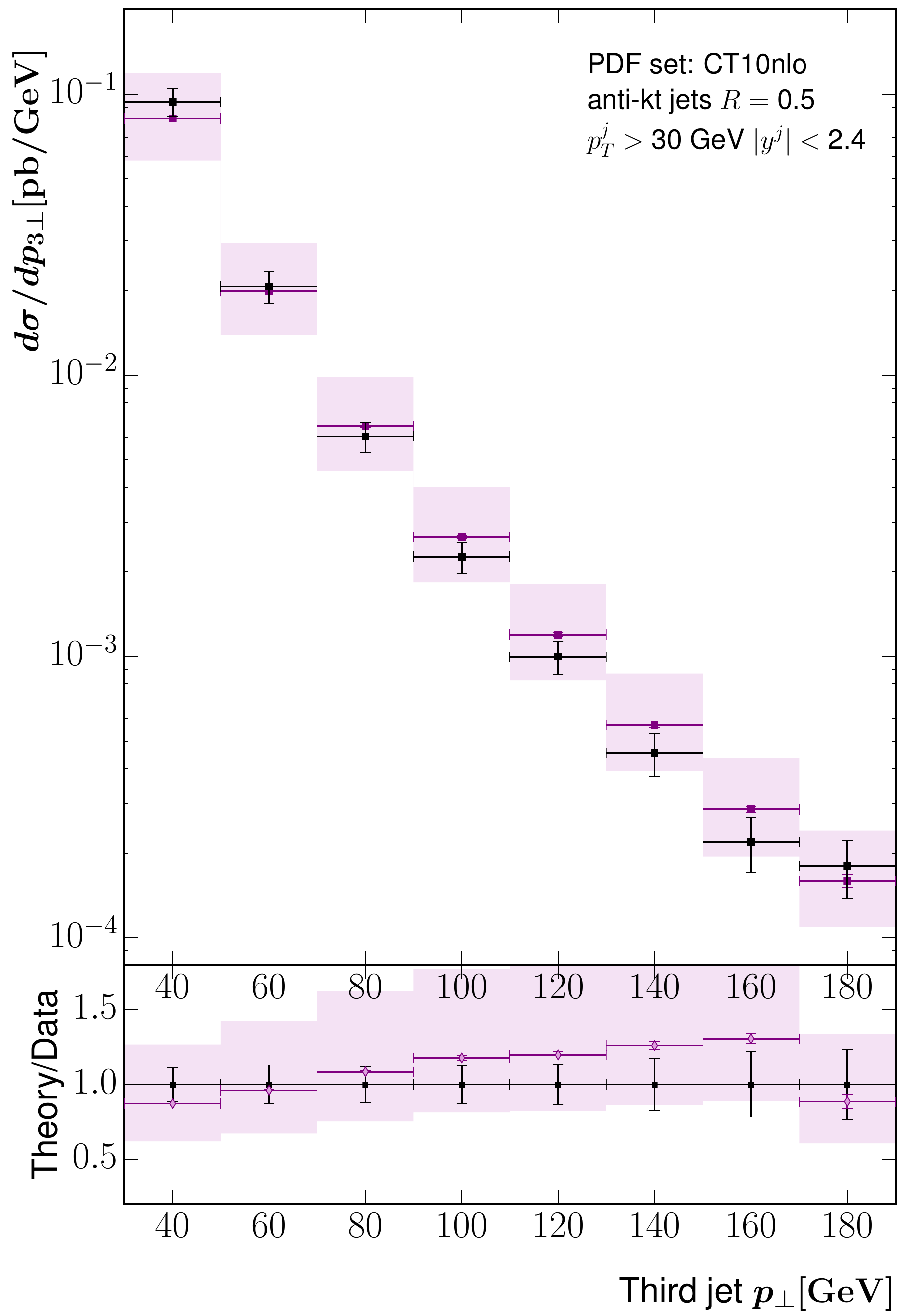}
    \caption{}
    \label{fig:HEJ_CMS_7b-2}
  \end{subfigure}
  ~
  \begin{subfigure}[b]{0.48\textwidth}
    \includegraphics[width=0.82\textwidth]{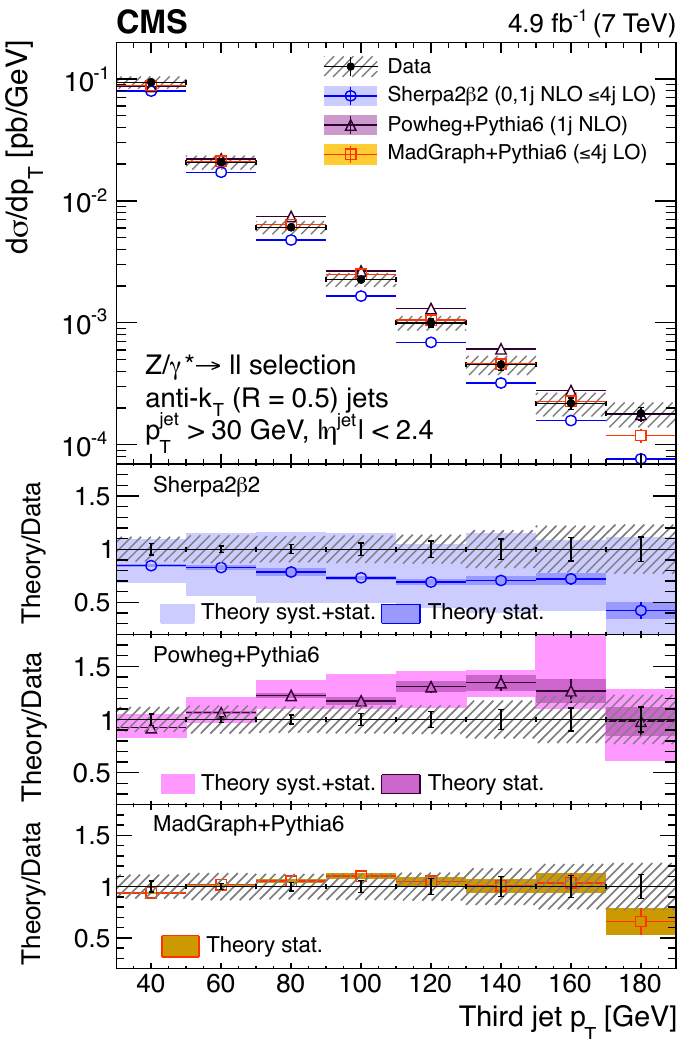}

    \vspace{0.1cm}
    \caption{}
    \label{fig:MC_CMS_7b-2}
  \end{subfigure}
  \caption{The transverse momentum distribution of the third hardest jet in
    inclusive dijet events in~\cite{Khachatryan:2014zya}, compared to (a) the
    predictions from HEJ and (b) the predictions from other theory descriptions.}
  \label{fig:CMS_3c}
  %
\end{figure}

We begin in Fig.~\ref{fig:CMS_2a} by showing the inclusive jet rates for these
cuts.  The HEJ predictions give a good description, especially for the 2- and
3-jet inclusive rates in this narrower phase space. In Figs.~\ref{fig:CMS_3b}--\ref{fig:CMS_3c}, we show the transverse momentum
distributions for the second and third jet respectively (the leading jet
distribution was not given for inclusive dijet events).  Beginning with the
second jet in Fig.~\ref{fig:CMS_3b}, we see that the HEJ predictions overshoot
the data at large transverse momentum.  In this region, the non-FKL matched
components of the HEJ description become more important and these are not
controlled by the high-energy resummation.  The HEJ predictions are broadly
similar to Powheg's $Z$-plus-one-jet NLO calculation matched with the Pythia
parton shower.  In contrast, Sherpa's central value significantly undershoots the
data at large transverse momentum although it is within their scale variation
band.  Here the Madgraph+Pythia central prediction gives the best
description of the data; their scale variation band is not shown.

Fig.~\ref{fig:CMS_3c} shows the transverse momentum distribution of the third
jet in this data sample.  Here, the ratio of the HEJ prediction to data shows a
linear increase with transverse momentum (until the last bin where all the
theory predictions show the same dip).  Both the Sherpa and Powheg central predictions
show similar deviations for this variable, although the data is just within the
larger Sherpa scale variation band. The Madgraph+Pythia prediction again
performs very well.


\clearpage
\subsection{Comparisons for the $W^\pm$+Jets/$Z$+Jets Ratio}
\label{sec:Ratios}

In this section we briefly comment on the all-order predictions from HEJ for the
ratio of $W^\pm$ plus jets to $\zg$ plus jets events.  We compare to data from a
recent study undertaken by the ATLAS collaboration \cite{Aad:2014rta}.  The cuts
for both final states are summarised in table \ref{tab:atlasWZcuts}.

\begin{table}[h]
  \centering
  \begin{tabular}{|l|l|}
    \hline
    Lepton Cuts & $p_{T\ell}>25$~GeV, \; $|\eta_\ell|<2.5$ \\
    &  $\Delta R^{\ell^+\ell^-} > 0.2$ \\ \hline
    Reconstructed $Z$ Cuts &  $66$~GeV $< m^{\ell^+\ell^-} <116$~GeV \\
    \hline
    Reconstructed $W^\pm$ Cuts & $m_{TW} > 40$~GeV\; $\slashed E_{T} > 25$~GeV \\ \hline
    Jet Cuts (anti-$k_T$, 0.4) & $p_{Tj}>30$~GeV, \; $|y_j|<4.4$ \\
    & $\Delta R^{j\ell} >0.5$ \\
\hline
  \end{tabular}
  \caption{Cuts applied to theory simulations in the analysis of the ATLAS $W^\pm$+jets/$Z$+jets ratio
    predictions shown in Tables \ref{table:RJetsIncl} and \ref{table:RJetsExcl}.}
  \label{tab:atlasWZcuts}
\end{table}

Tables~\ref{table:RJetsIncl} and \ref{table:RJetsExcl} show the measured values
of the ratio between $W$-plus-jets and $Z$-plus-jets events, $R_{jet}$,
separated into inclusive and exclusive samples of 2, 3 and 4 jets.  Also shown
are the corresponding values from HEJ and the ratio of the two.  We see
extremely good agreement for the 2-jet ratios and the 3- and 4-jet ratios agree at
the 10\% level.  This is comparable with the other
theoretical predictions used in the study
(\texttt{BlackHat+SHERPA}~\cite{Berger:2009ep,Berger:2010vm,Berger:2010zx,Ita:2011wn},
\texttt{ALPGEN}~\cite{Mangano:2002ea} and
\texttt{SHERPA}~\cite{Gleisberg:2008ta,Hoeche:2012yf}) as can be seen in Fig.~1
of \cite{Aad:2014rta}.

  \begin{table}[!h]
  \begin{center}
  \begin{tabular}{| c | c | c | c |}
          \hline
  $N_{jets}$ & Data & HEJ & HEJ/Data \\ \hline
  $\ge2$ & $8.64\pm0.04(\mbox{stat.})\pm0.33(\mbox{syst.})$ & $8.66\pm0.12(\mbox{stat.})^{+0.14}_{-0.16}(\mbox{s.v.})$ & $1.00\pm0.01(\mbox{stat})^{+0.02}_{-0.01}(\mbox{s.v.})$ \\ \hline
  $\ge3$ & $8.18\pm0.08(\mbox{stat.})\pm0.52(\mbox{syst.})$ & $7.96\pm0.25(\mbox{stat.})^{+0.01}_{-0.01}(\mbox{s.v.})$ & $0.97\pm0.03(\mbox{stat})^{+0.01}_{-0.00}(\mbox{s.v.})$ \\ \hline
  $\ge4$ & $7.62\pm0.20(\mbox{stat.})\pm0.95(\mbox{syst.})$ & $8.55\pm0.69(\mbox{stat.})^{+0.02}_{-0.02}(\mbox{s.v.})$ & $1.12\pm0.09(\mbox{stat})^{+0.00}_{-0.00}(\mbox{s.v.})$ \\ \hline
  \end{tabular}
  \caption{The HEJ prediction for inclusive $R_{jet}$ rates at 2, 3 and 4 jets compared with ATLAS data.}
  \label{table:RJetsIncl}
  \end{center}
  \end{table}

  \begin{table}[!h]
  \begin{center}
  \begin{tabular}{| c | c | c | c |}
          \hline
  $N_{jets}$ & Data & HEJ & HEJ/Data \\ \hline
  2 & $8.76\pm0.05(\mbox{stat.})\pm0.31(\mbox{syst.})$ & $8.88\pm0.135(\mbox{stat.})^{+0.15}_{-0.18}(\mbox{s.v.})$ & $1.01\pm0.02(\mbox{stat})^{+0.021}_{-0.02}(\mbox{s.v.})$ \\ \hline
  3 & $8.33\pm0.10(\mbox{stat.})\pm0.45(\mbox{syst.})$ & $7.85\pm0.265(\mbox{stat.})^{+0.01}_{-0.01}(\mbox{s.v.})$ & $0.94\pm0.01(\mbox{stat})^{+0.001}_{-0.03}(\mbox{s.v.})$ \\ \hline
  4 & $7.69\pm0.21(\mbox{stat.})\pm0.71(\mbox{syst.})$ & $8.44\pm0.684(\mbox{stat.})^{+0.04}_{-0.04}(\mbox{s.v.})$ & $1.10\pm0.01(\mbox{stat})^{+0.005}_{-0.09}(\mbox{s.v.})$ \\ \hline
  \end{tabular}
  \caption{The HEJ prediction for exclusive $R_{jet}$ rates at 2, 3 and 4 jets compared with ATLAS data.}
  \label{table:RJetsExcl}
  \end{center}
  \end{table}


\section{Conclusions}
\label{Conclusions}

In this paper we have discussed augmenting the theoretical description of
inclusive $\zg$-plus-dijets processes with the dominant logarithms in the High
Energy limit at all orders in $\alpha_s$.  In particular, the description
constructed here is accurate to leading logarithm in $\hat{s}/\hat{t}$.  This is
achieved within the High Energy Jets (HEJ) framework.  We began in
section~\ref{sec:HEQCD} by motivating and describing the construction of an
approximation to the hard-scattering matrix element for an arbitrary number of
gluons in the final state.  This uses factorised currents for electroweak boson
emission and outer jet production combined with a series of (gauge-invariant)
effective vertices for extra QCD real emissions.

In contrast to previous HEJ constructions (for pure jets, $W$-plus-jets and
Higgs boson-plus-jets), the complete description of the interference
contributions between $Z$ and $\gamma^*$ processes \emph{and} between forward
and backward emissions required a new regularisation procedure.  This is
described in section~\ref{sec:Constructing} where we showed explicitly the
cancellation of real and virtual divergences by using the Lipatov ansatz to
include the dominant contributions in the High Energy limit of the all-order
virtual contributions.  The method by which we match our matrix element to the
leading order matrix elements was also outlined here. In this way we
achieve the formal accuracy of our Monte Carlo predictions to Leading
Logarithmic in $(\hat{s}/\hat{t})$ and merge Leading Order predictions in $\alpha_s$ for the
production of two, three or four jets.

In section~\ref{sec:Comparisons}, we compared the predictions of our
construction to $\zg$-plus-jets data collected at the ATLAS and CMS experiments
during Run I.  We see excellent agreement for a wide range of observables and
can be seen to describe regions of phase space well where some other
fixed-order-based predictions do not fare as well.  Discrepancies which occur
only do so in regions where we do not expect this description to perform as
well, for example where there is a large ratio between $p_{T1}$ and $p_{T2}$.
We also discuss properties of other available theoretical descriptions.

This all-order description of $\zg$-plus-dijets allows predictions for the ratio
of $W^\pm$+dijets to $\zg$+dijets at all-orders in $\alpha_s$ for the first
time.  This is an extremely important analysis as many theoretical and
experimental uncertainties cancel in this ratio and in section~\ref{sec:Ratios},
we show that we correctly reproduce the ratios of the total cross sections.

Just as for previous analyses of LHC data, it is found that the high-energy logarithms
contained in HEJ are necessary for a satisfactory description of data in key regions of phases
space, e.g.~at large values of jet invariant mass. Such regions of phase
space are crucial for the analysis of Higgs boson production in association
with dijets. The impact of the high-energy 
logarithms will only be more pronounced at the larger centre-of-mass energy of
LHC Run II, and beyond at a possible future circular collider.  The HEJ
framework and Monte
Carlo is the unique flexible event generator to contain these corrections and
will provide important theoretical input for the study of important processes
at LHC Run II and beyond.  

\section*{Acknowledgements}
\label{sec:acknowledgements}

This research was supported in part by the Research Executive Agency (REA) of
the European Union under PITN-GA-2012-315877 (MCnetITN).  JRA and JJM are
supported by the UK Science and Technology Facilities Council (STFC).  JMS is supported
by a Royal Society University Research Fellowship.


\appendix
\boldmath
\section{Dependence on the Regularisation Parameter, $\lambda_{cut}$}
\unboldmath
\label{sec:indep-lambd}

In this appendix, we show results for various values of the parameter
$\lambda_{cut}$ defined in section~\ref{sec:Constructing}.  We
increase our sensitivity to the parameter by showing results for FKL momentum
configurations only.  The non-FKL samples which are added to give the total
cross sections have no dependence on $\lambda_{cut}$ and would therefore dilute
any dependence in the full sample.  We begin in table~\ref{tab:lambdaxs} where we show the value of
the cross section for different values of $\lambda_{cut}$ for exclusive 2-, 3-
and 4-jet samples.  The cuts applied are the same as in section \ref{sub:ATLAS}.
It is clear that the cross section does not display a large dependence on the
value of $\lambda_{cut}$.

\begin{table}[htp!]
\begin{center}
\begin{tabular}{| c | c | c | c |}
\hline
$\lambda_{cut}$ (GeV) & $\sigma(2j)$ ($pb$) & $\sigma(3j)$ ($pb$) & $\sigma(4j)$ ($pb$) \\ \hline
0.2 & $5.03 \pm 0.02$ & $0.70 \pm 0.02$ & $0.13 \pm 0.03$ \\
0.5 & $5.05 \pm 0.01$ & $0.70 \pm 0.01$ & $0.13 \pm 0.01$ \\
1.0 & $5.09 \pm 0.01$ & $0.71 \pm 0.01$ & $0.13 \pm 0.01$ \\
2.0 & $5.16 \pm 0.04$ & $0.72 \pm 0.01$ & $0.13 \pm 0.01$ \\ \hline
\end{tabular}
\caption{The FKL-only cross sections for the 2-, 3- and 4-jet exclusive rates with associated statistical errors shown for different values of the regularisation parameter
$\lambda_{cut}$.  The scale choice was half the sum over all transverse scales in the event, $H_T/2$.}
\label{tab:lambdaxs}
\end{center}
\end{table}

\begin{figure}[htp!]
  \centering
  \begin{subfigure}[b]{0.3\textwidth}
    \includegraphics[width=\textwidth]{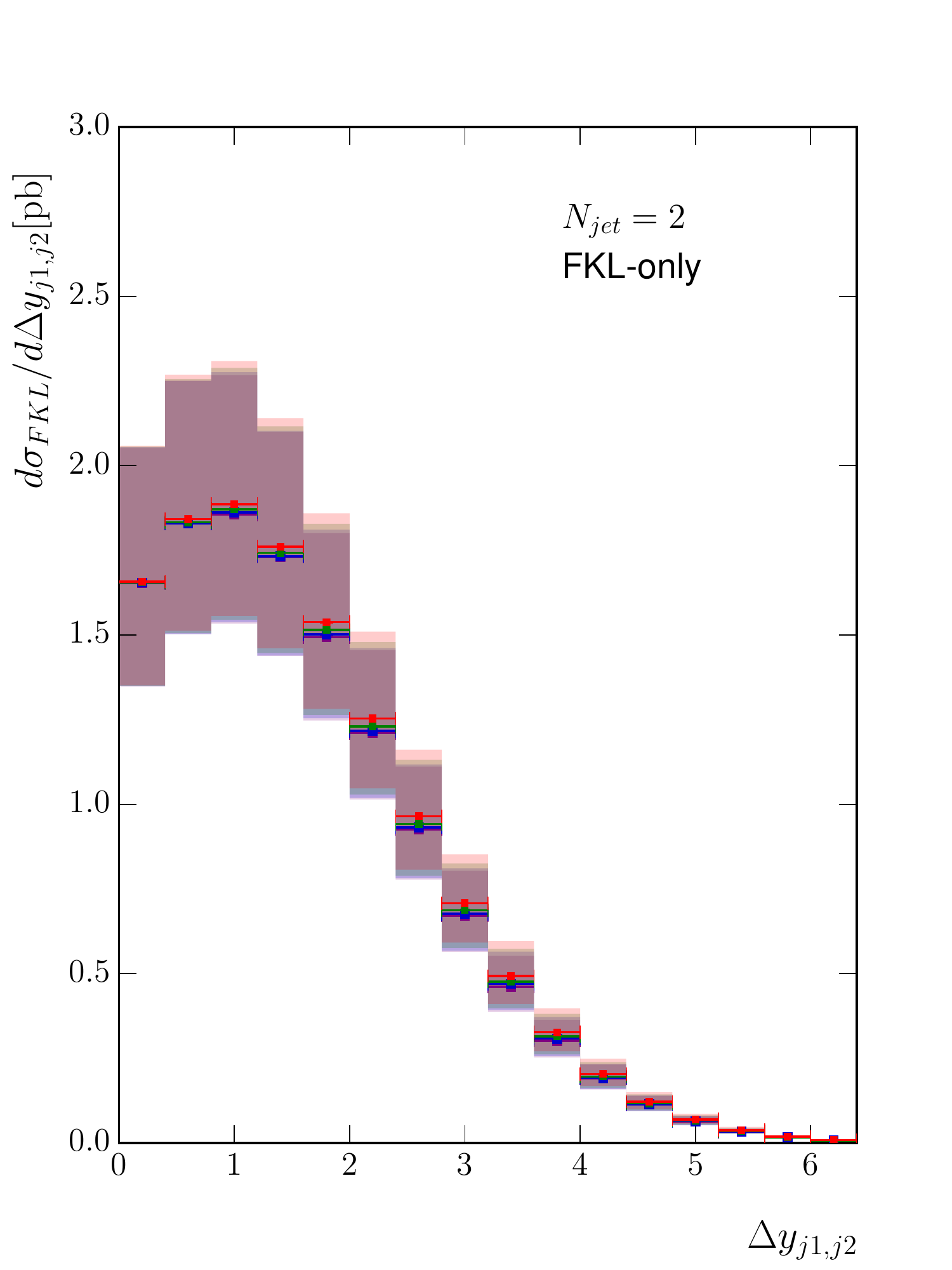}
    \caption{}
    \label{fig:lambdadist1}
  \end{subfigure}
  \begin{subfigure}[b]{0.3\textwidth}
    \includegraphics[width=\textwidth]{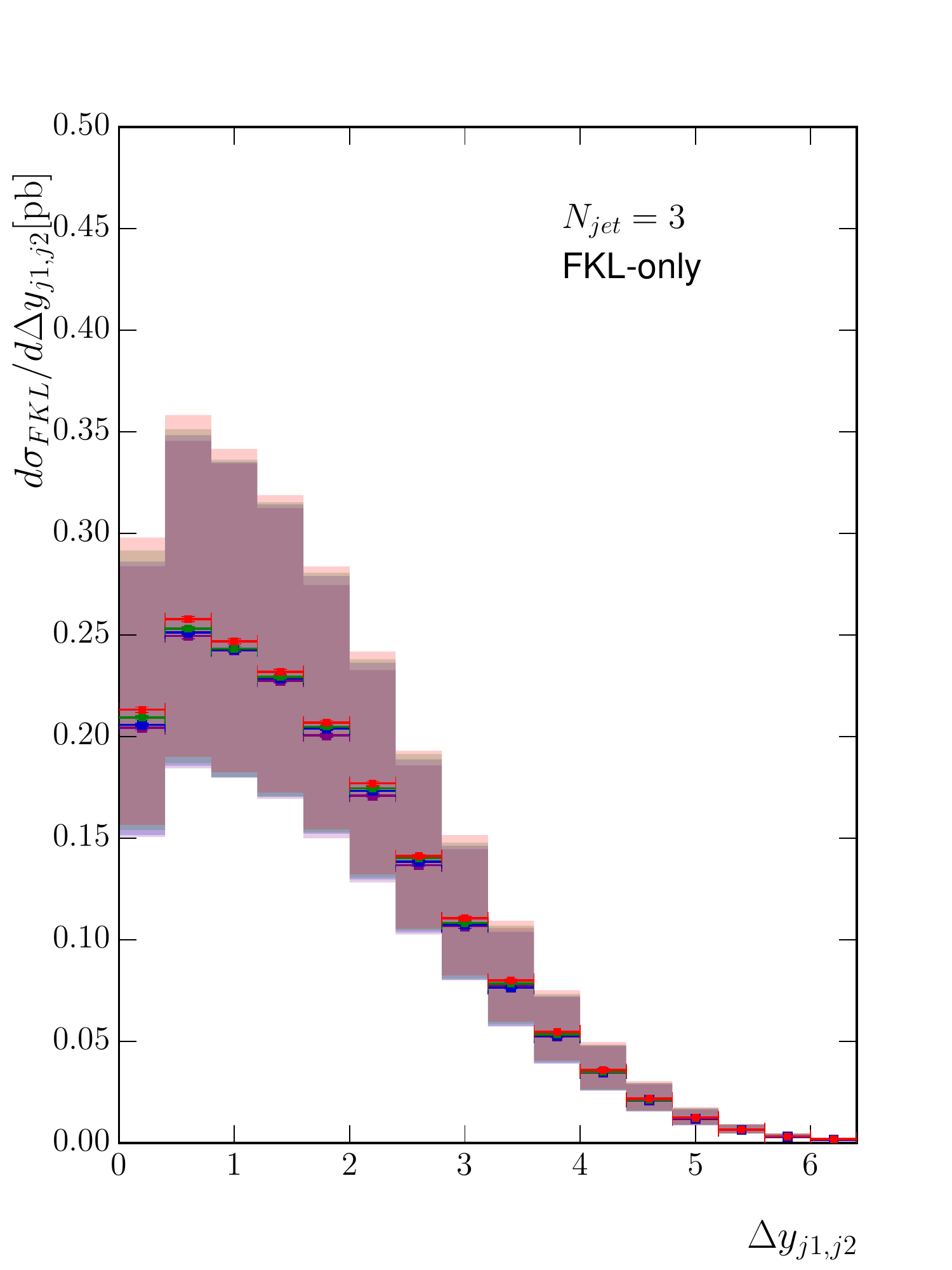}
    \caption{}
    \label{fig:lambdadist2}
  \end{subfigure}
  \begin{subfigure}[b]{0.3\textwidth}
    \includegraphics[width=\textwidth]{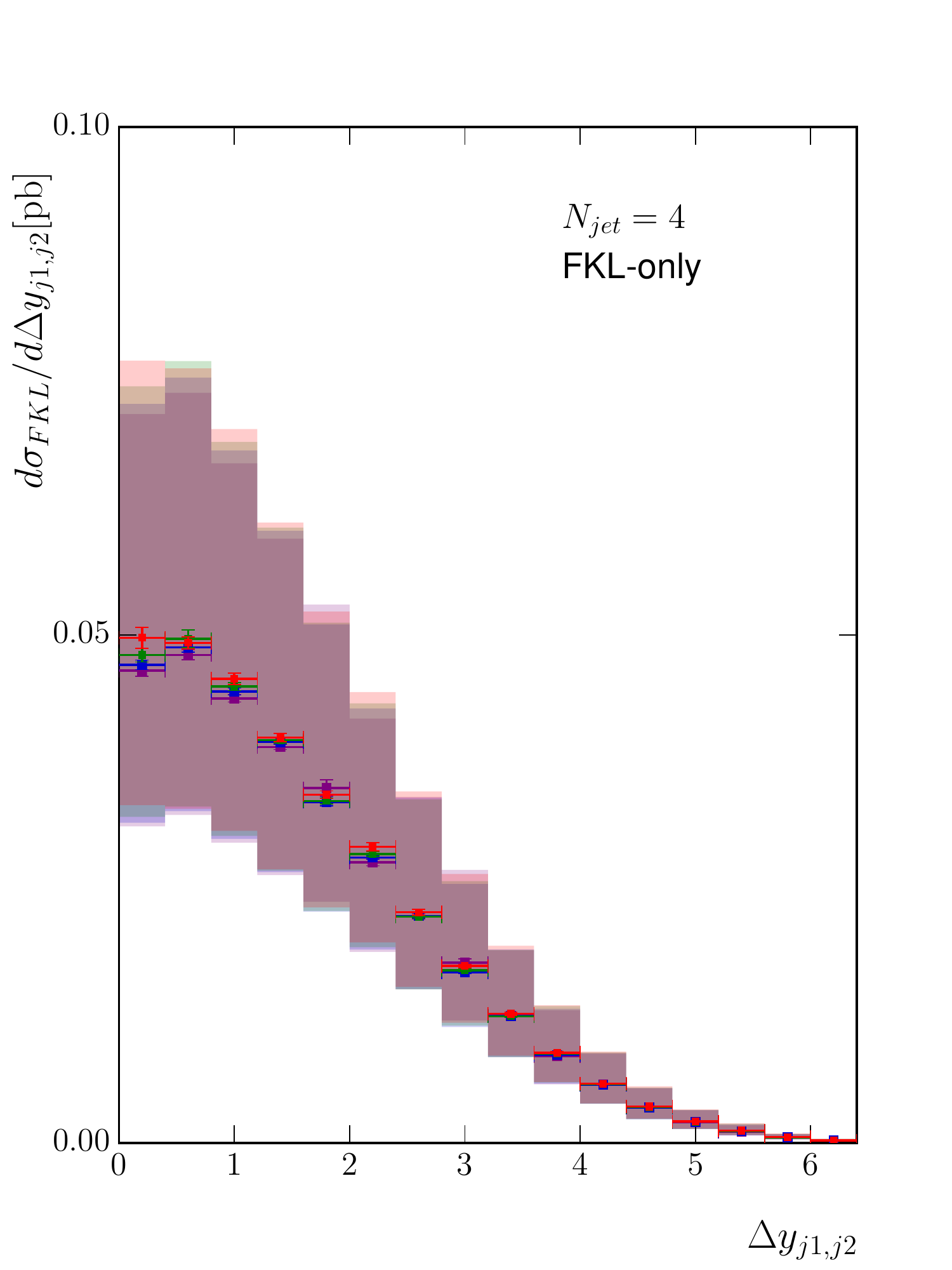}
    \caption{}
    \label{fig:lambdadist3}
  \end{subfigure}

  \begin{subfigure}[b]{0.3\textwidth}
    \includegraphics[width=\textwidth]{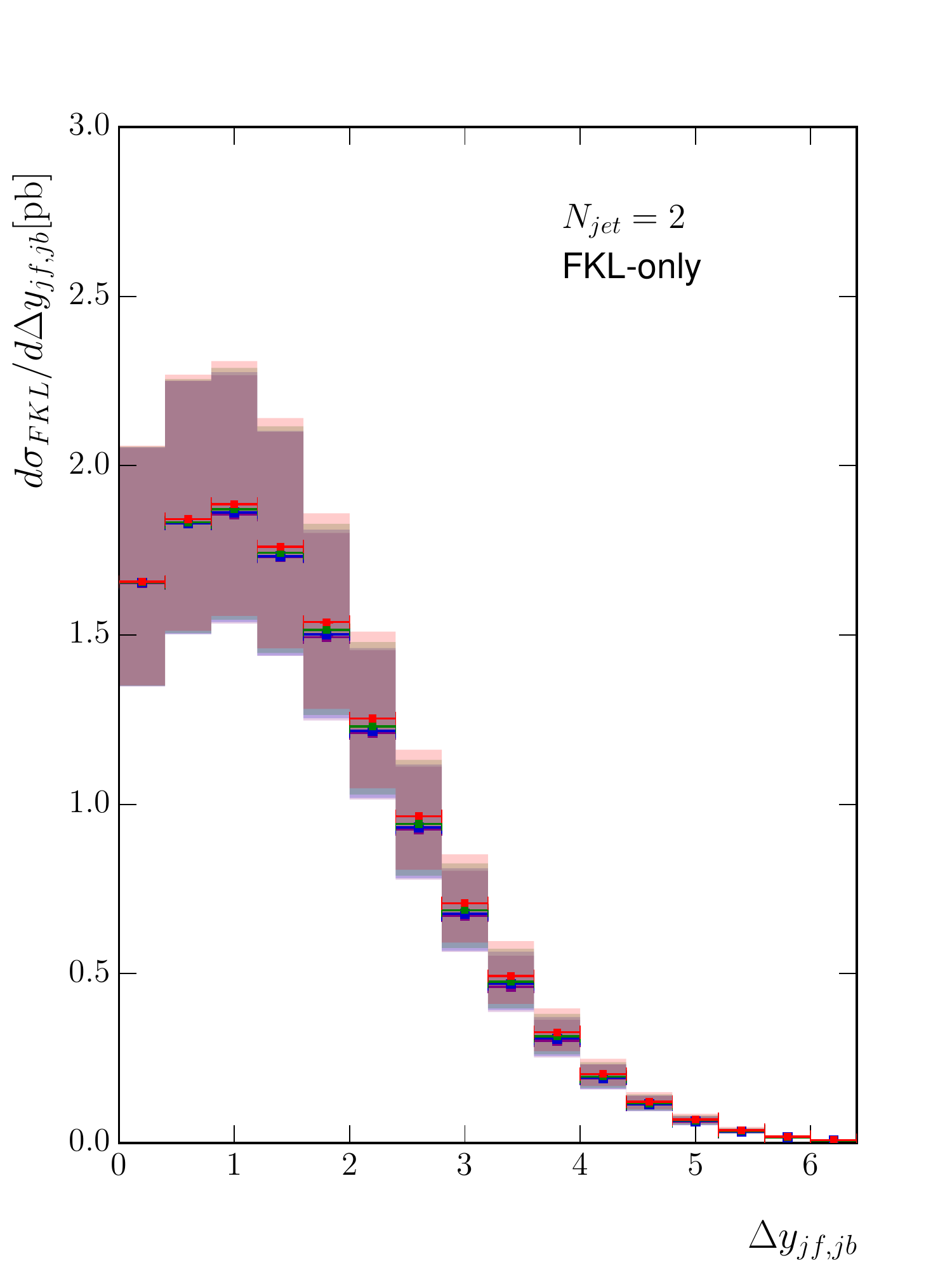}
    \caption{}
    \label{fig:lambdadist4}
  \end{subfigure}
  \begin{subfigure}[b]{0.3\textwidth}
    \includegraphics[width=\textwidth]{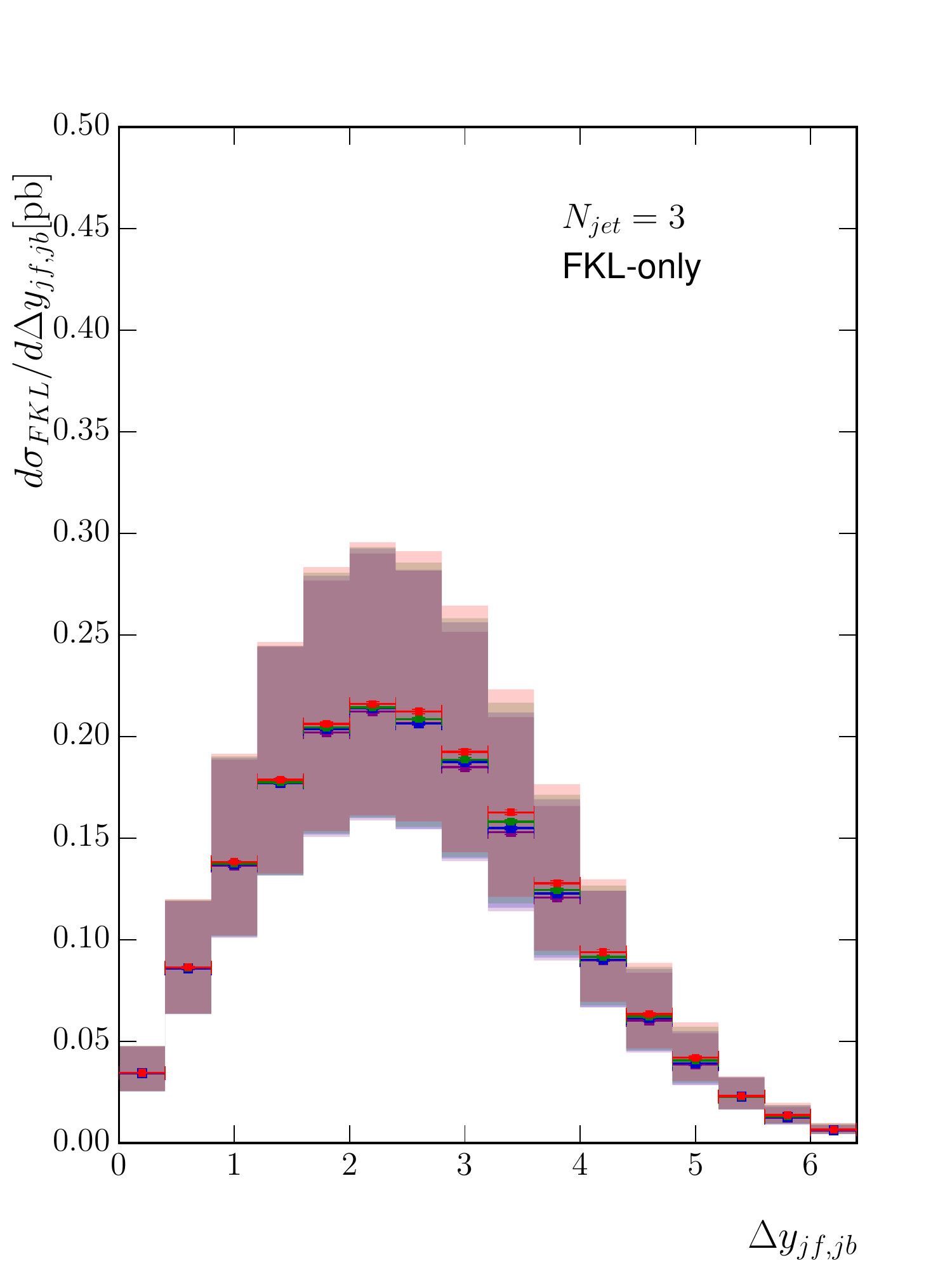}
    \caption{}
    \label{fig:lambdadist5}
  \end{subfigure}
  \begin{subfigure}[b]{0.3\textwidth}
    \includegraphics[width=\textwidth]{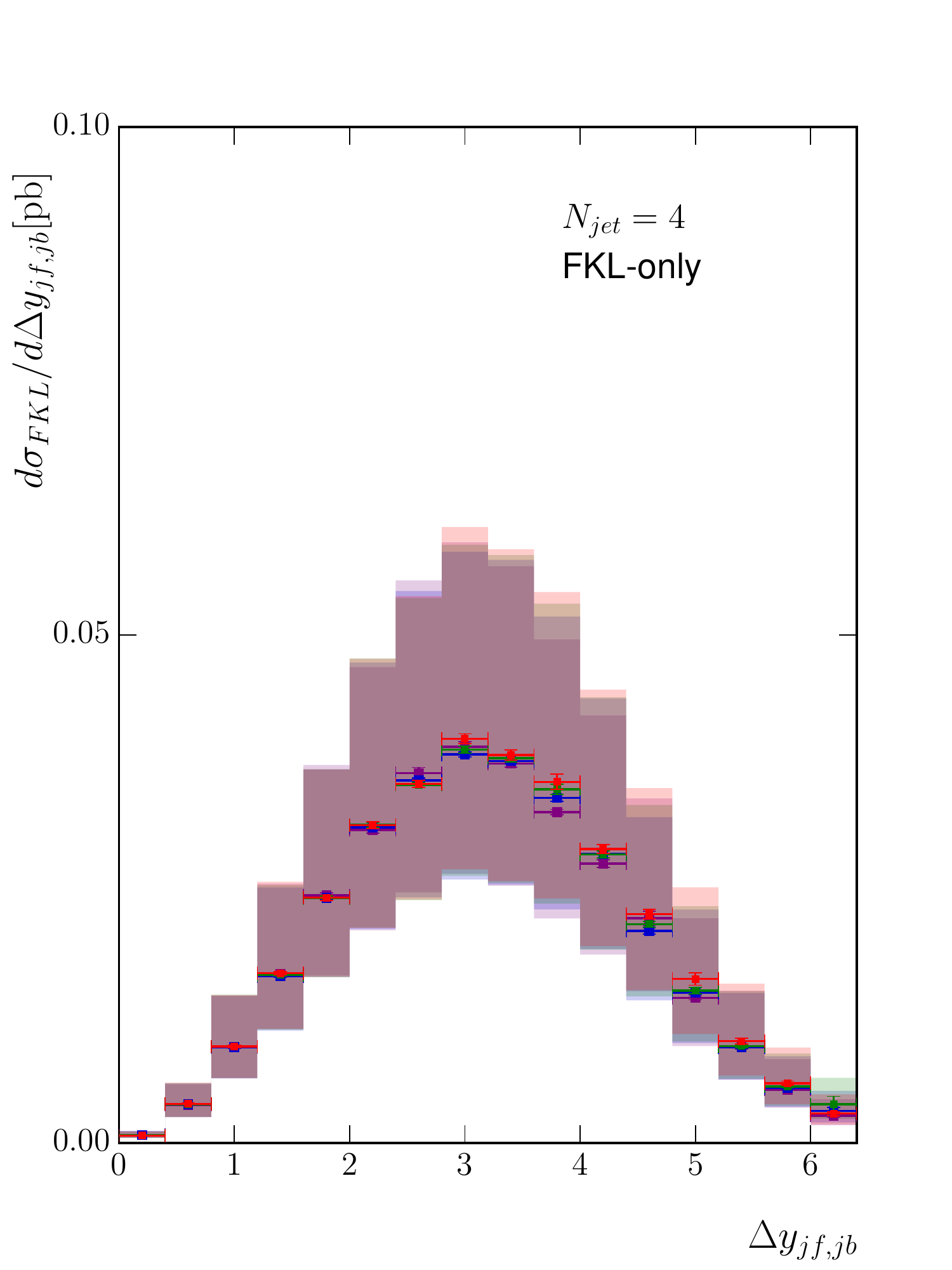}
    \caption{}
    \label{fig:lambdadist6}
  \end{subfigure}

  \caption{(a)--(c) The effect of varying $\lambda_{cut}$ on the differential distribution
in the rapidity gap between the two leading jets in $p_\perp$, $\Delta y_{j1, j2}$, with the $N_{jet}=2,3,4$
exclusive selections shown from left to right,
and (d)--(f) for the rapidity gap between the most extremal jets in rapidity, $\Delta y_{jf, jb}$, with the $N_{jet}=2,3,4$
exclusive selections shown from left to right.
The different colours represent $\lambda_{cut}=0.2$ (red), 0.5 (blue), 1.0
(green) and 2.0 (purple) and the bands represent the scale variation described in
the text.}
  \label{fig:lambdadist}

\end{figure}

Figure~\ref{fig:lambdadist} shows the effect of the same variation in $\lambda_{cut}$ on the
differential distribution in both the rapidity gap between the two leading jets in $p_\perp$,
$\Delta y_{j1, j2}$, (a)--(c), and the rapidity gap between the two extremal jets in
rapidity, $\Delta y_{jf, jb}$, (d)--(f).  Results are shown for exclusive 2-, 3-
and 4-jet samples in each case, once again the cuts applied are the same as in section \ref{sub:ATLAS}.
Again the scale choice for the central line was $\mu_F=\mu_R=H_T/2$.  The variation bands
have been determined by varying these two scales independently by up to a factor of two
in either direction with the extremal points removed where the relative difference between
$\mu_F$ and $\mu_R$ is greater than a factor of 2.  The distributions also show a very weak
dependence on the choice of $\lambda_{cut}$.

In practice, our default chosen value for $\lambda_{cut}$ is 0.2.

\boldmath
\section{Normalisation Effects on Scale Uncertainties in $\zg$+Jets}
\unboldmath
\label{sec:normalisation}

\begin{figure}[btp]

  \centering

  \begin{subfigure}[b]{0.35\textwidth}
    \includegraphics[width=0.96\textwidth]{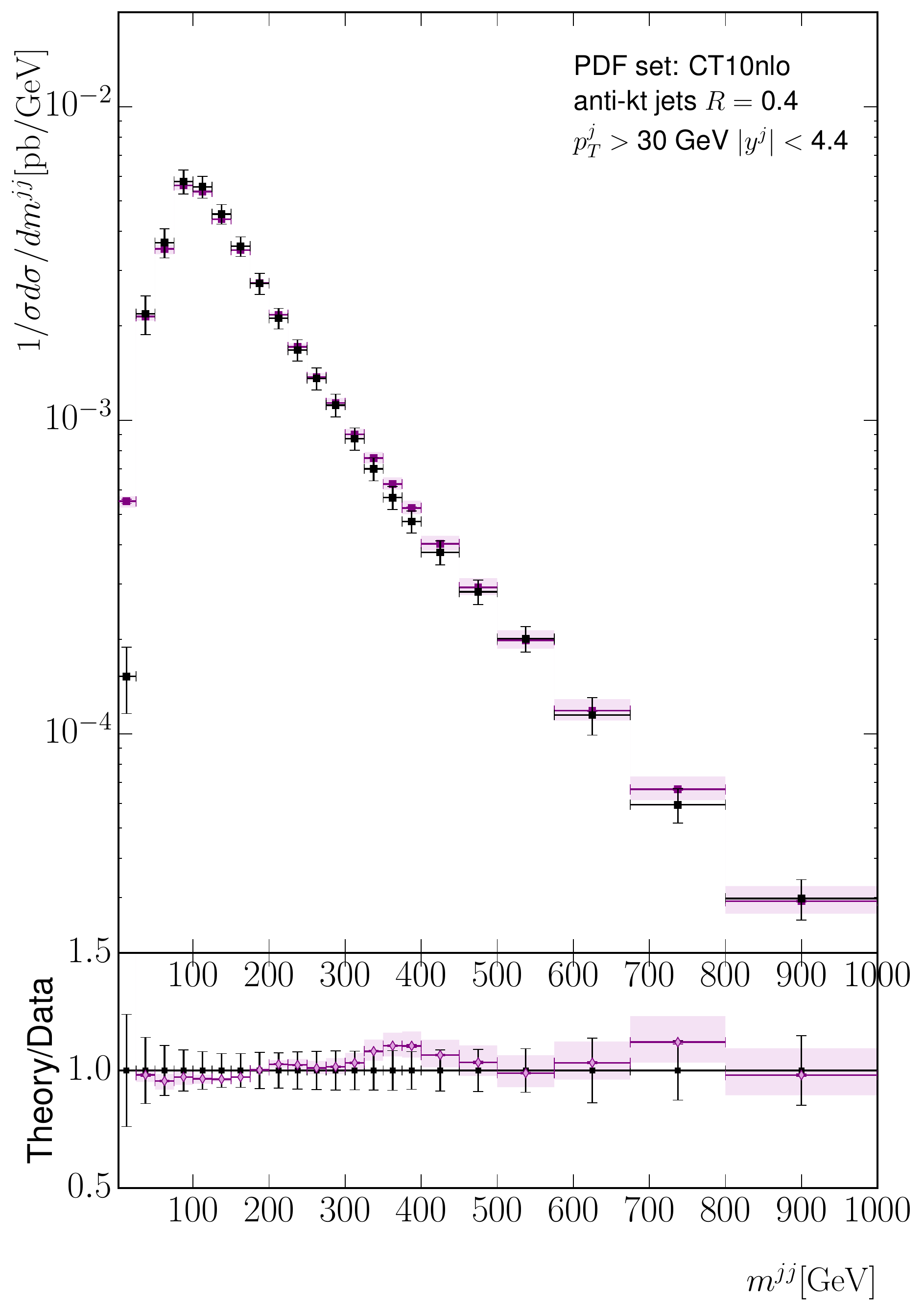}
    \caption{}
    \label{fig:ATLAS_Z_11b_norm}
  \end{subfigure}
  ~
  \begin{subfigure}[b]{0.35\textwidth}
    \includegraphics[width=0.9\textwidth]{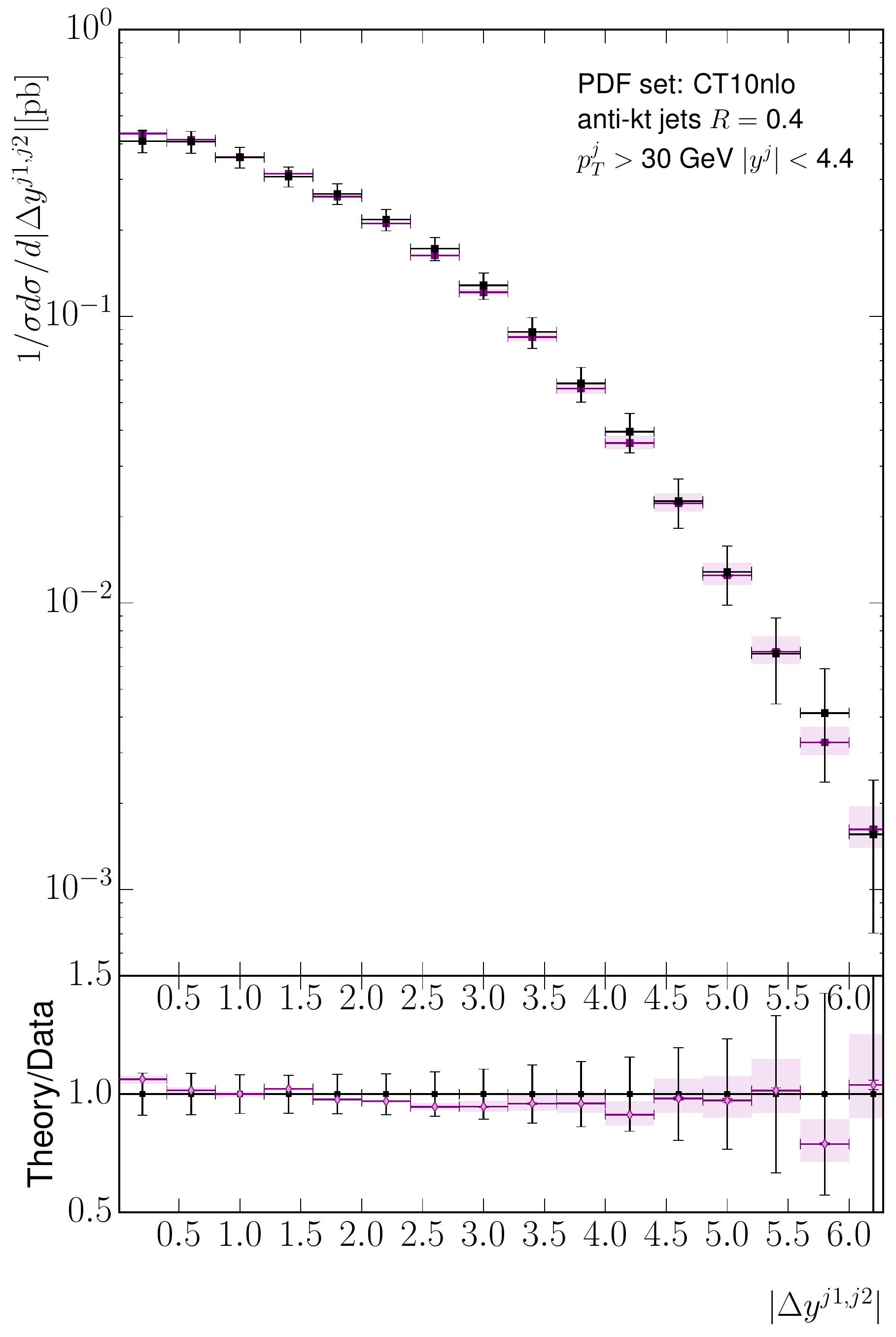}
    \vspace{0.2cm}
    \caption{}
    \label{fig:ATLAS_Z_11a_norm}
  \end{subfigure}

  \begin{subfigure}[b]{0.35\textwidth}
    \includegraphics[width=0.9\textwidth]{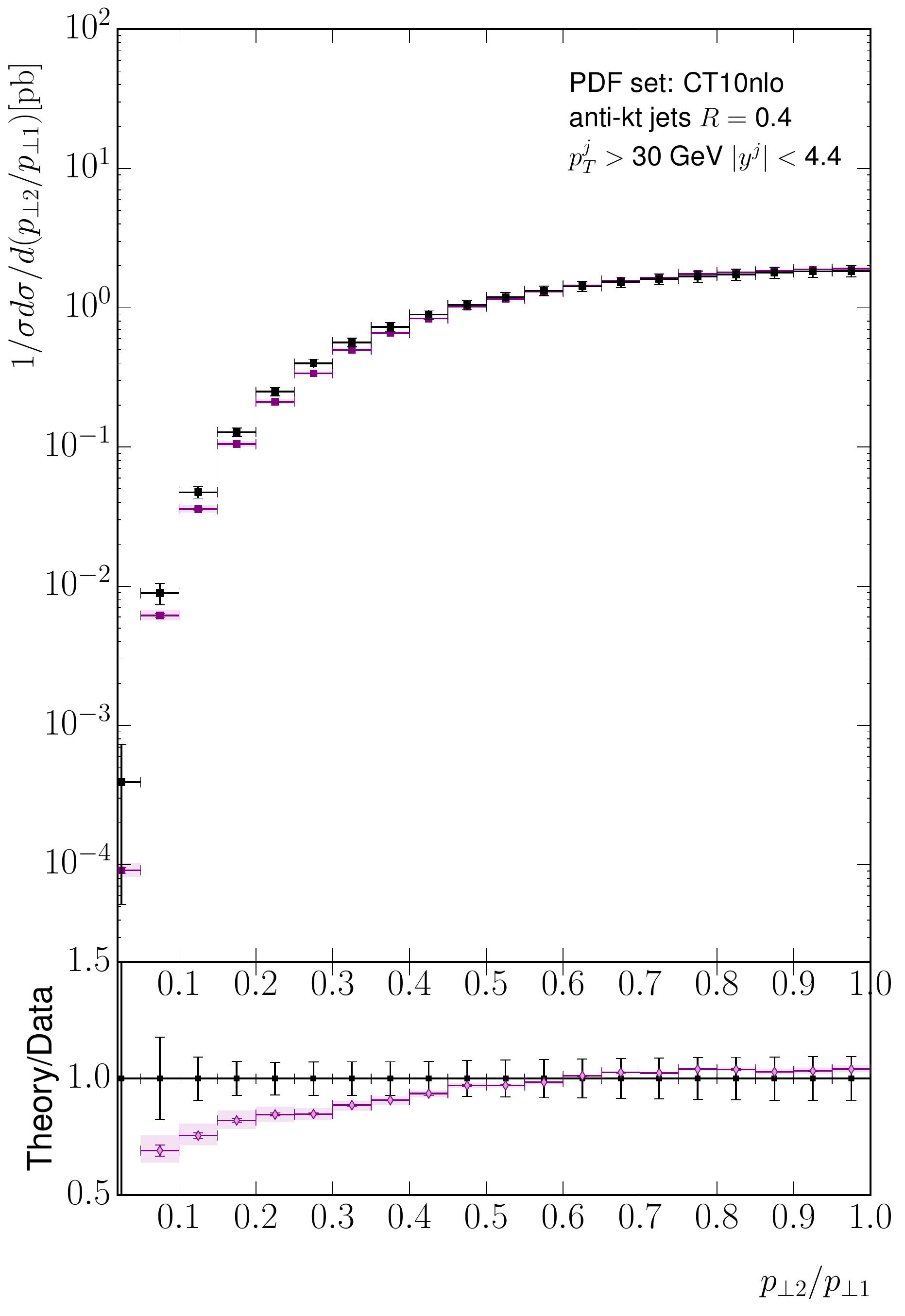}
    \vspace{0.2cm}
    \caption{}
    \label{fig:ATLAS_Z_7b_norm}
  \end{subfigure}
  \caption{The predictions of section~\ref{sub:ATLAS} normalised to the total
    cross-section, with scale variation consistently applied to numerator and denominator.}

  \label{fig:ATLAS_norm}
\end{figure}

Here we discuss the effect of normalising the predictions shown in section~\ref{sub:ATLAS}
to the total cross-section.  We see from Fig.~\ref{fig:HEJ_ATLAS_2a} that we describe
the experimentally observed inclusive two jet rate very well and, as such, do not require
normalisation to agree with the data.  However, applying a normalisation
procedure which consistently applies scale variation simultaneously in numerator
and denominator significantly reduces the size of the scale uncertainty bands
for High Energy Jets (or any
theoretical prediction).

In Figs.~\ref{fig:ATLAS_Z_11b_norm},~\ref{fig:ATLAS_Z_11a_norm} and~\ref{fig:ATLAS_Z_7b_norm}
we show the results from Figs.~\ref{fig:HEJ_ATLAS_11b},~\ref{fig:HEJ_ATLAS_11a} and~\ref{fig:HEJ_ATLAS_7b}
where we have normalised to the total cross-section calculated for each scale
combination.  We see that, as expected, the central value of HEJ still describes the data
well in the regions discussed in section~\ref{sub:ATLAS} and now the size of the
theoretical uncertainty band is significantly reduced (by as much as a factor of 16 in the last bin of the $p_{\perp 2}/p_{\perp 1}$-distribution for
example, and more typically by a factor of about 4).  This illustrates that
varying the renormalisation and factorisation scales leads to a change in overall normalisation but not
to any significant change in shape.  Therefore, it is still valuable to discuss
the quality of agreement of the central line, despite their apparently large accompanying
uncertainty bands in the unnormalised predictions.
\newpage


\bibliographystyle{JHEP}
\bibliography{ZJetsPaper}
\end{document}